\documentclass[lettersize,journal]{IEEEtran}
\usepackage{amsmath,amsfonts}
\usepackage{algorithm}  
\usepackage{algpseudocode}  
\usepackage{graphics}
\usepackage{textcomp}
\usepackage{amsmath} 
\usepackage{mathtools}
\usepackage{amssymb}  
\usepackage{color}
\usepackage{url}
\usepackage{commath}
\usepackage{array} 
\usepackage{color} 
\usepackage[caption=false]{subfig}
\usepackage{footnote}
\makesavenoteenv{tabular}
\makesavenoteenv{table}
\makesavenoteenv{figure}

\usepackage{siunitx}

\usepackage{tikz}
\usepackage{booktabs}
\usepackage{multirow}
\usepackage{graphicx}
\usepackage{adjustbox}
\usepackage{comment}
\usepackage{dblfloatfix}

\usepackage{hyperref}
\hypersetup{
    colorlinks=true,
    linkcolor=blue,
    filecolor=magenta,      
    urlcolor=cyan,
    }
    
\usepackage{color}
\usepackage{colortbl}

\usepackage{hyperref}

\title{\LARGE \bf Husformer: A Multi-Modal Transformer for \\ Multi-Modal Human State Recognition}

\author{Ruiqi Wang$^{1}\dag$, Wonse Jo$^{1}\dag$, Dezhong Zhao$^{2}$, Weizheng Wang$^{1}$\\ Baijian Yang$^{1}$, Guohua Chen$^{2}$, and Byung-Cheol Min$^{1}$

\thanks{$^{1}$Department of Computer and Information Technology, Purdue University, West Lafayette, IN 47907, USA \tt\small[{wang5357, jow, wang5716, byang, minb] @purdue.edu}}
\thanks{$^{2}$College of Mechanical and Electrical Engineering, Beijing University of Chemical Technology, Beijing, China. \tt\small{DZ\_Zhao@buct.edu.cn}, {chengh@mail.buct.edu.cn}}
\thanks{$\dag$ Equal contribution}
}

\begin{document}
\maketitle
\captionsetup{font={small}}

\begin{abstract}

Human state recognition is a critical topic with pervasive and important applications in human-machine systems. Multi-modal fusion, which involves combining metrics from multiple data sources, has been shown to be an effective method for improving the recognition performance. While promising results have been reported by recent multi-modal-based models, they generally fail to leverage the sophisticated fusion strategies that would model sufficient cross-modal dependencies when producing the fusion representation; instead, they rely on costly and inconsistent feature crafting and alignment. To address this limitation, we propose an end-to-end multi-modal transformer framework for multi-modal human state recognition called \textit{Husformer}. Specifically, we propose using cross-modal transformers, which inspire one modality to reinforce itself through directly attending to latent relevance revealed in other modalities, to fuse different modalities while ensuring sufficient awareness of the cross-modal interactions introduced. Subsequently, we utilize a self-attention transformer to further prioritize contextual information in the fusion representation. Extensive experiments on two human emotion corpora (DEAP and WESAD) and two cognitive load datasets (MOCAS and CogLoad) demonstrate that in the recognition of human state, our \textit{Husformer} outperforms both state-of-the-art multi-modal baselines and the use of a single modality by a large margin, especially when dealing with raw multi-modal features. We also conducted an ablation study to show the benefits of each component in \textit{Husformer}\footnote{~Source code for our \textit{Husformer} and experiments is available at: \url{https://github.com/SMARTlab-Purdue/Husformer}.}.

\end{abstract}


\section{Introduction}
\label{sec:introduction}
\IEEEPARstart{R}{ecognition} of human state, including human affective state, known as emotion, and cognitive load, known as mental stress, plays an enormous role in any human-machine interaction systems, as enabling machines to perceive, understand, and adapt to different human emotional and cognitive states, and improves the performance of the whole systems \cite{zhang2014recognition,wickens2002multiple,yang2019assessing}. Human state assessment methods can generally be divided into two main categories based on the types of signals used: physiological and behavioral assessments. Physiological assessments involve measuring human physiological metrics, such as galvanic skin response (GSR), electroencephalography (EEG), electrooculography (EOG), electrocardiography (ECG), electromyogram (EMG), and heart rate (HR), which change in response to involuntary reactions of the human nervous system under specific states. On the other hand, behavioral assessments analyze subconscious human behavioral responses, including facial expressions, body and eye movements, and mouse movements, and associate them with different human states \cite{marshall2002index}.

Unfortunately, owing to the complexity of human state reasoning, it is improbable that signals from a single modality are sufficient to achieve optimal recognition performance in terms of accuracy and robustness \cite{debie2019multimodal,sebe2005multimodal,he2020advances}. For instance, different signals possess different levels of sensitivity to different task environments and even different human subjects, and determining one modality that is efficient for every task scenario and subject is impractical and impossible. Furthermore, reliance on a unimodal data source means that noise or interruption of the signals can result in extensive errors or even a failure of the recognition system \cite{brooks1998multi}.

Recently, multi-modal fusion-based human state recognition methods that combine data from multiple modalities have been proposed and demonstrated promise as a solution for the aforementioned challenges faced by single-modal-based approaches \cite{chen2016robust,kruger2018multimodal,huang2017fusion,tang2017multimodal}. The adoption of multi-modal signals can reduce the noise-to-signal ratio and enhance tolerance against sensor failures. More importantly, fusing different metrics collected from the same subject under one particular human state through multiple modalities can reveal important and comprehensive indexes of human emotion and cognitive load that are inaccessible via a single modality \cite{debie2019multimodal,he2020advances}. Nevertheless, the inherent heterogeneity of multiple modalities poses challenges in generating an efficient fusion index of the human state. These challenges include: 1) different modalities are typically unaligned, resulting in unbalanced feature length and temporal resolutions; and 2) combining multi-modal features may lead to the inclusion of biased or irrelevant information due to feature noncommensurability across multiple modalities;  and 3) inference of long-term and complicated dependencies across modalities is required for accurate fusion \cite{debie2019multimodal}.

\begin{figure*}[t]
    \centering
    \includegraphics[width=0.9\linewidth]{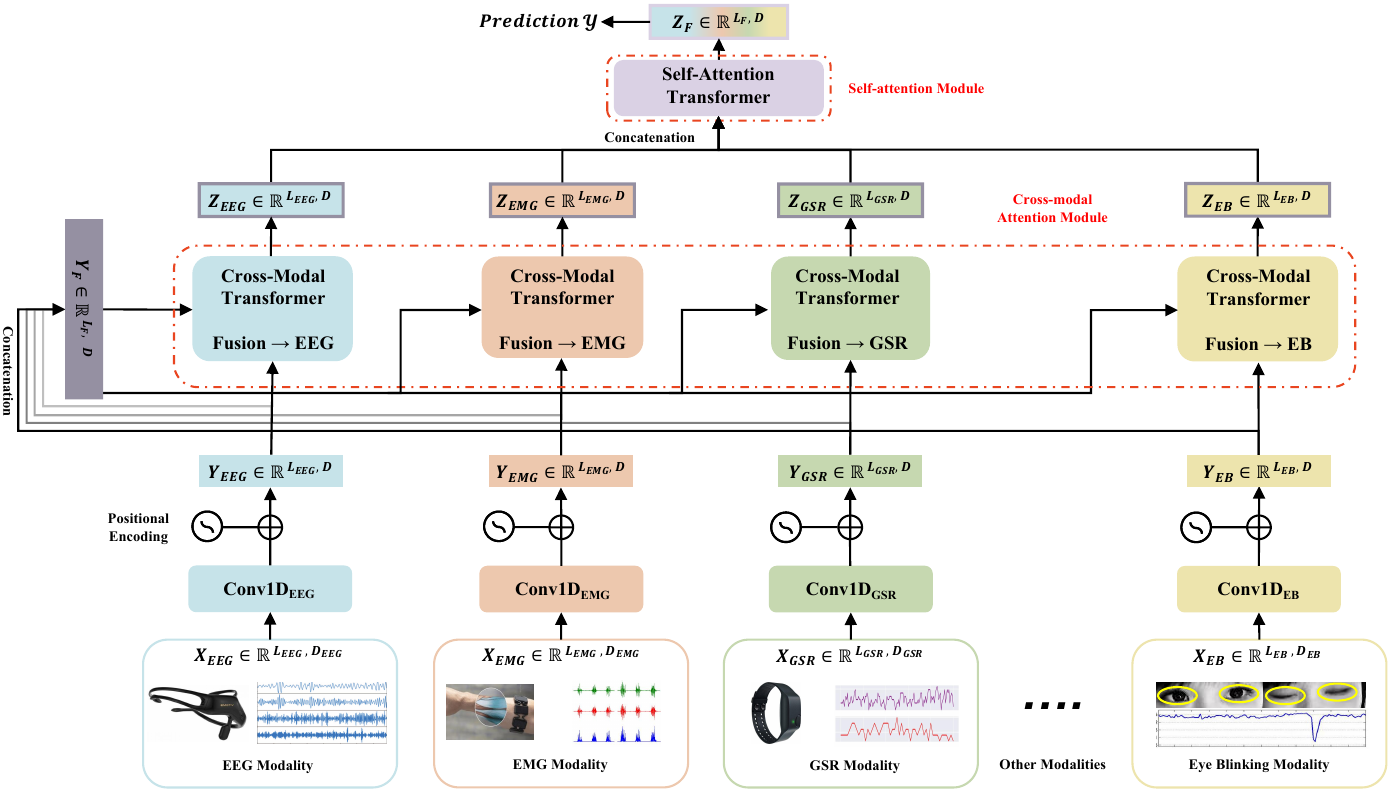}
    \caption{Framework of the proposed \textit{Husformer}, taking multi-modal data input from four example modalities: EEG, EMG, GSR, and eye blinking (EB). The multi-modal data inputs, $X_{(.)} \in \mathbb{R}^{L_{(.)},D_{(.)}}$, where $L_{(.)}$ and $D_{(.)}$ separately present the length and dimension of the input sequence of one modality, are passed through multiple one-dimension temporal convolution layers, $Conv1D_{(.)}$ (Section \ref{TC}), and then encoded with positional information to produce the low-level unimodal features $Y_{(.)} \in \mathbb{R}^{L_{(.)},D}$ which all have the same dimension $D$ (Section \ref{PE}); these are concatenated to generate the low-level fusion representation $Y_{F} \in \mathbb{R}^{{L_{F}},D}$. This representation is then fed alongside the unimodal low-level features of each modality into each respective cross-modal attention transformer, wherein the target modality is adapted and reinforced according to the other resource modalities through learning the attention between its unimodal features and the low-level representation (Section \ref{CM}). Then all reinforced unimodal features $Z_{(.)} \in \mathbb{R}^{L_{(.)},D}$ are concatenated into the mid-level fusion representation, which is passed through a self-attention transformer to generate the high-level fusion representation $Z_{F} \in \mathbb{R}^{L_{F},D}$ with important contextual information prioritized. Finally, the high-level fusion representation is transported to fully-connected layers to make predictions (Section \ref{SM} and \ref{MT}). }
    \label{fig:framework}
\end{figure*}

Current multi-modal fusion approaches for human state recognition remain in their early stages and have not yet fully addressed the challenges arising from the heterogeneity across multiple modalities. Most methods rely on extensive feature engineering and alignment to concatenate features from different modalities and produce the fusion representation\cite{debie2019multimodal,singh2022multimodal,horii2018modeling}. However, such direct concatenation fusion schemes ignore the latent correlations across modalities and may be limited by non-instantaneous coupling, even if manually aligned. While several methods have been proposed to learn cross-modal interactions by training shared representations \cite{tang2017multimodal,zheng2018emotionmeter,qiu2018multi,liu2021comparing},  their network structures are too shallow to capture complicated cross-modal dependencies and introduce sufficient complementarity across different modalities. Additionally, the feature crafting methods required before fusion necessitated by these approaches are costly and may differ for different modalities, and their parameters require prior expert knowledge and extensive cross-validation to be optimal \cite{zhou2021cognitive,zhou2020multimodal}. Such expensive and inconsistent feature engineering undermines the simplicity of the model and its applicability to new task scenarios. Furthermore, to our best knowledge, none of the existing multi-modal methods have been shown to effectively predict both human affective state and cognitive load. Most methods are tailored for a specific combination of modalities under certain task scenarios, leading to a potential deficiency of generality.


To address the aforementioned gaps, we present \textit{Husformer}: an end-to-end multi-modal transformer for human state recognition that efficiently learns representations of human state from heterogeneous multi-modal streams. Figure \ref{fig:framework} illustrates its structure with four example input modalities: EEG, EMG, GSR, and eye blinking (EB). The core components of \textit{Husformer} are the cross-modal attention module and self-attention module, which consist of multiple cross-modal attention transformers and one self-attention transformer, respectively. The cross-modal attention transformers model the latent interactions across modalities by continuously adapting and reinforcing features from one modality with those of other modalities (e.g., EEG $\leftarrow$ EMG, GSR, and EB). Unlike direct concatenation of multi-modal modalities or learning cross-modal shared representations through shallow neural networks, our cross-modal attention mechanism encourages the target modality to directly attend to low-level features in other modalities where strongly relevant and complementary information is revealed. This leads to more adaptive and efficient complementarity and cooperation across multiple modalities. The self-attention transformer prioritizes important contextual information in the fusion representation concatenated from reinforced unimodal features of all modalities supplied by the cross-modal attention transformers. This finally generates a weighted high-level fusion representation based on which predictions are made.

To evaluate the performance of \textit{Husformer}, we conducted extensive experiments on four multi-modal datasets: DEAP \cite{koelstra2011deap} and WESAD \cite{schmidt2018introducing} for emotion recognition, and MOCAS \cite{wonse_jo_2022_7023242} and CogLoad \cite{gjoreski2020datasets} for cognitive load estimation. Additionally, We performed a comprehensive ablation study to investigate the benefits of each module in \textit{Husformer}.

The main contributions of this work can be summarized as follows: 
\begin{itemize}
    \item \textit{Husformer} is an end-to-end model that learns directly and efficiently from heterogeneous multi-modal physiological and behavioral signals without the massive feature crafting and alignment required in previous works.
    \item We introduce cross-modal attention transformers to fuse features from different modalities and sufficiently model long-term cross-modal interactions, and then a self-attention transformer to prioritize effectual contextual information in the fusion representation. 
    \item To the best of our knowledge, this is the first time a generic model for human state recognition is presented and proven to be effective for both human emotion and cognitive load.
    \item Our extensive experiments on four publicly available datasets demonstrate the benefits of the \textit{Husformer} and each of its constituent modules.
\end{itemize}

\section{Background}
\label{sec:related_work}

This section reviews existing research on human state recognition using multi-modal fusion approaches and the preliminary transformer networks that serve as the basis of our model. 

\subsection{Multi-Modal Fusion for Human State Recognition}
Given the complex nature of human state, a single modality is insufficient to achieve recognition with satisfactory performance in terms of accuracy and robustness, especially in real-world task scenarios where signals are more subject to interruption, noise, and delay \cite{he2020advances}. To solve this issue, multi-modal fusion that integrates human signals from more than one source modality into a synchronized compact representation has been adopted. Nevertheless, multi-modal fusion methods for human state recognition are still at the initiatory stages and suffer from several defects in fusion strategies. Firstly, the direct concatenation of different modalities at sensor, feature, or decision level has been adopted for most existing works to produce the fusion representation \cite{debie2019multimodal,singh2022multimodal}. This fusion strategy may result in meaningless added information or even bias and noise owing to the feature noncommensurability across different modalities. For instance, simply fusing HR with other modalities can introduce additional valuable information in some task scenarios but bring disruption in others, as HR-related signals are not only influenced by cognitive load and emotion but also by irrelevant physical activities \cite{debie2019multimodal}. Also, directly fusing different modalities whose features have unbalanced length and temporal dimensions necessitates considerable feature alignment preprocessing, which could reduce the richness and diversity of the input data, potentially limiting the model's ability to capture complex patterns and relationships within or across modalities. In general, simple concatenation operations fail to model the cross-modal interactions and thus are unlikely to capture important representations that would otherwise be revealed by multi-modal fusion.

Lately, some advanced fusion schemes have been proposed to solve the above issues; these aim to model cross-modal interactions by training shared representations across different modalities. For example, \cite{tang2017multimodal} adopts restricted Boltzmann machines (RBM) \cite{ngiam2011multimodal} to train a hidden layer that is expected to learn the shared representations of sub-layers from different modalities; \cite{tsai2019multimodal} introduces cross-modal temporal correlations obtained via learning shared weights across sub-layers of different modalities; and \cite{qiu2018multi} utilizes deep canonical correlation analysis (DCCA) \cite{andrew2013deep} to maximize the correlations among features of two modalities. However, due to their utilization of shallow network structures, these existing methods are unlikely to comprehensively capture tangled correlations across different modalities, especially from raw multi-modal signals where features are more interrupted. Moreover, the DCCA utilized in \cite{qiu2018multi,liu2021comparing} can only analyze the correlation between two modalities, limiting the application of the model to general task scenarios that may require the simultaneous fusion of more than two modalities. 

Additionally, existing works usually require extensive feature crafting procedures to reach sound recognition performance due to the abundant collinearity or heterogeneity in the raw multi-modal features. In addition, different modalities are usually preprocessed by different methods whose optimal parameters are initially unknown. For example, in \cite{zhou2018multimodal}, infinite impulse response
(IIR) high-pass and Hanning window filters were used for EEG modality feature engineering while continuous decomposition analysis and other physiological libraries were utilized for GSR and HR modalities. Afterward, independent component analysis (ICA) and manual feature selection procedures depending on prior knowledge were adopted to preprocess candidate features a second time. Such involved non-uniform feature engineering procedures harm the simplicity and universality of the model, making it difficult to apply to general task scenarios and especially to realistic applications, which runs counter to the original intention of multi-modal fusion. In addition, as far as we know, existing works only focus on the recognition of either affective state or cognitive state, and most are tailored for specific task scenarios and not validated on publicly available datasets; thus, the models lack generality and replicability.

Distinct from these prior studies, our proposed \textit{Husformer} 1) focuses on the general recognition of human state, including emotion and cognitive load; 2) does not require any feature alignment and extensive feature crafting procedures in existing works, but rather learns from raw multi-modal feature streams; and 3) utilizes cross-modal attention transformers as the fusion strategy, thereby introducing efficient cooperation and complementary adaptions across modalities regardless of the number of modalities.

\subsection{Transformer Network}
\label{TN}
The transformer network was originally proposed by \cite{vaswani2017attention} to solve sequence-to-sequence machine translation tasks in the natural language processing area. Unlike the traditional encoder-to-decoder structure, the transformer network adopts a multi-head self-attention mechanism to substitute for the attention-based convolution and recurrence layers. The self-attention mechanism aims to calculate a global representation of a sequence input that reveals meaningful contextual information by relating different components within the sequence. Basically, a self-attention block adapts each entity in a sequence by considering the global contextual information of the whole sequence. Multi-head self-attention splits the attention into multiple latent sub-spaces (heads), which enables the modeling of multiple complex contextual relations across elements in the sequence, leading to a more comprehensive global representation.

\section{Approach}
\label{sec:approach}
In this section, we present \textit{Husformer}, an end-to-end multi-modal transformer for multi-modal human affective and cognitive state recognition that learns the fusion representation directly and efficiently from raw multi-modal data streams.

\subsection{Overview}
As depicted in Fig. \ref{fig:framework}, from a high-level perspective, the \textit{Husformer} uses a position-wise feed-forward process to fuse multi-modal signal series supplied by  multiple cross-modal transformers (Section \ref{CM}). In each cross-modal transformer, the target modality is repeatedly enhanced with low-level features from other modalities by calculating the latent cross-modal attention between the target low-level unimodal feature and the low-level fusion representation. A sequence-to-sequence model, i.e., a self-attention transformer, is then utilized to process the mid-level fusion representation sequence consisting of all enhanced unimodal features and generate the adaptively weighted high-level fusion representation (Section \ref{SM}). Specifically, by computing multi-head self-attention (Section \ref{TN}), the self-attention transformer analyzes the pairwise relationships across elements in the mid-level fusion representation, namely, the reinforced unimodal features, to calculate adaptive weights at different positions and so highlight critical contextual information. Finally, the high-level fusion representation is passed through fully-connected layers to make predictions (Section \ref{MT}).

\subsection{Temporal Convolutions}
\label{TC}
Let $M_1$, $M_2$, $...$ $M_n$ denote $n$ modalities. And let $X_{\{M_1,..., M_n\}} \in \mathbb{R}^{L_{M_1,..., M_n}, D_{M_1,..., M_n}}$ present the raw multi-modal data sequences input from these $n$ modalities. $L_{(.)}$ and $D_{(.)}$ represent the sequence length (e.g., channel number) and dimension (e.g., sampling rate) of each unimodal input respectively in this paper. The multi-modal input sequences are passed through multiple one-dimension temporal convolution layers with different kernels to generate multiple convoluted sequences $\dot{X}_{\{M_1,..., M_n\}} \in \mathbb{R}^{L_{{M_1,..., M_n}},D}$ with the same dimension $D$:
\begin{equation}
\dot{X}_{\{M_1,..., M_n\}} = \operatorname{Conv1D}\left(X_{\{M_1,..., M_n\}}, k_{\{M_1,..., M_n\}}\right)
\label{eqconv}
\end{equation}
\noindent where, $k_{\{M_1,..., M_n\}}$ denotes the temporal convolution kernel sizes for $n$ modalities $M_1$, $M_2$, $...$ $M_n$. 

Each convoluted sequence aims to contain low-level temporal features of each modality. Furthermore, it is important that after temporal convolutions, different unimodal input sequences are projected to the same dimension, making the dot-product calculation in the following cross-modal attention module mathematically feasible.

\subsection{Positional Encoding}
\label{PE}
As mentioned in the introduction of the transformer network (Section \ref{TN}), the transformer model has no inherent awareness of the positional information of each sequence component, such as the relative or absolute position of features within a modality sequence. To introduce sufficient awareness of relations across neighboring elements, i.e., features of adjacent channels within one modality sequence, and thus spatial information, we follow the method proposed in \cite{vaswani2017attention} and apply positional encoding (PE) to the convoluted sequences $\dot{X}_{\{M_1,..., M_n\}}$ using $sin$ and $cos$ functions with different frequencies. The PE of one convoluted sequence $\dot{X} \in \mathbb{R}^{L,D}$ can be defined as a matrix:
\begin{equation}
\begin{aligned}
\mathrm{PE_{\dot{X}}}[pos, 2k] &=\sin \left(\frac{pos}{10000^{\frac{2 k}{D}}}\right) \\
\mathrm{PE_{\dot{X}}}[pos, 2k+1] &=\cos \left(\frac{pos}{10000^{\frac{2 k}{D}}}\right)
\end{aligned}
\end{equation}
\noindent where $pos  \in [1, ..., L]$ and $k \in [0, ..., \frac{D}{2})$.

Each characteristic dimension (i.e., column) of $PE_{\dot{X}}$ is a position index displayed in the sinusoidal pattern. The calculated PEs $PE_{\dot{X}_{\{M_1,..., M_n\}}} \in \mathbb{R}^{L_{{M_1,..., M_n}},D}$, are then augmented with convoluted sequences $\dot{X}_{\{M_1,..., M_n\}}$ to obtain low-level unimodal feature sequences with both initial temporal and spatial information encoded $Y_{{M_1,..., M_n}} \in \mathbb{R}^{L_{{M_1,..., M_n}},D}$:
\begin{equation}
Y_{\{M_1,..., M_n\}} =\dot{X}_{\{M_1,..., M_n\}}+PE_{\dot{X}_{\{M_1,..., M_n\}}}
\label{eq_pe}
\end{equation}

Additionally, the extracted low-level unimodal feature sequences of all modalities are then concatenated to produce the low-level fusion representation $Y_{F} \in \mathbb{R}^{{L_{F}},D}$:
\begin{equation}
{Y_{F}} = \operatorname{Concat}\left(Y_{M_1}, \ldots, Y_{M_n}\right)
\label{eq_fus}
\end{equation}




\subsection{Cross-modal Attention Module}
\label{CM}

To provide sufficient complementary interactions and adaptions across different modalities, we respectively feed the low-level unimodal feature sequence of each modality $Y_{M_i}\in \mathbb{R}^{L_{M_i},D}$ with the low-level multi-modal fusion representation $Y_{F} \in \mathbb{R}^{{L_{F}},D}$ to a cross-modal attention module that is comprised of multiple cross-modal transformer networks. Each cross-modal transformer is expected to continuously reinforce low-level unimodal features of the target modality with those of other source modalities by learning cross-modal attention between the input unimodal sequence and the low-level fusion representation. Namely, the learned cross-modal attention inspires each target modality to directly engage in low-level unimodal features of other source modalities, which are encoded in the low-level fusion representation, to adaptively find relevant and useful information that can serve as complementary reinforcements for itself. In addition to fostering adaptive and accurate awareness of correlations across multiple modalities, this module can also help address potential artifacts and feature incompatibilities across within raw multi-modal features by disregarding irrelevant aspects, such as interrupted or insensitive features in the low-level unimodal feature sequence, when calculating cross-modal attention.

\subsubsection{Cross-modal Attention}

\begin{figure}[h]
\centering
\subfloat[Illustration of the cross-modal attention between each low-level unimodal feature and the low-level fusion representation.\label{fig:cm}]{\includegraphics[width=\linewidth]{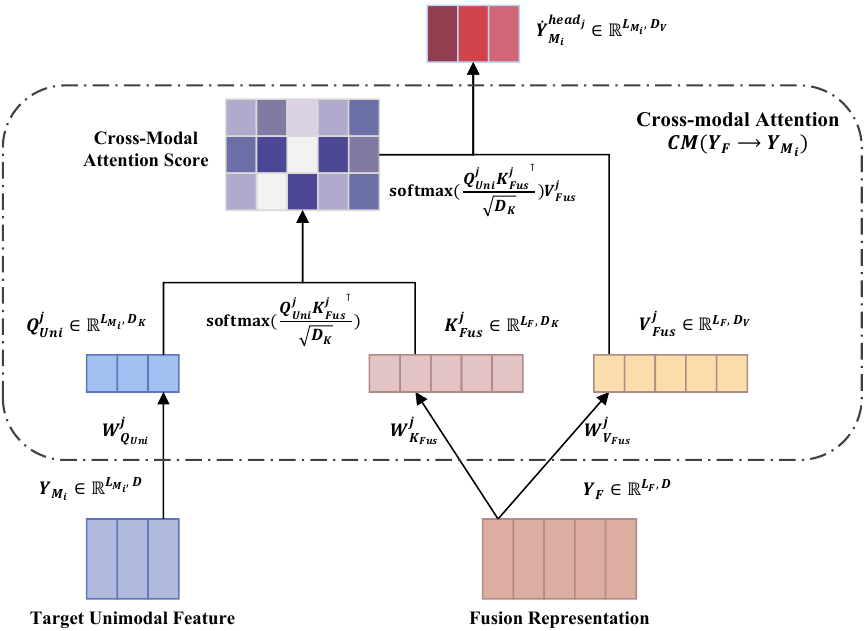}}

\subfloat[Illustration of the cross-modal Transformer network stacked by multiple identical cross-modal attention encoder layers.\label{fig:cmt}
]{\includegraphics[width=0.9\columnwidth]{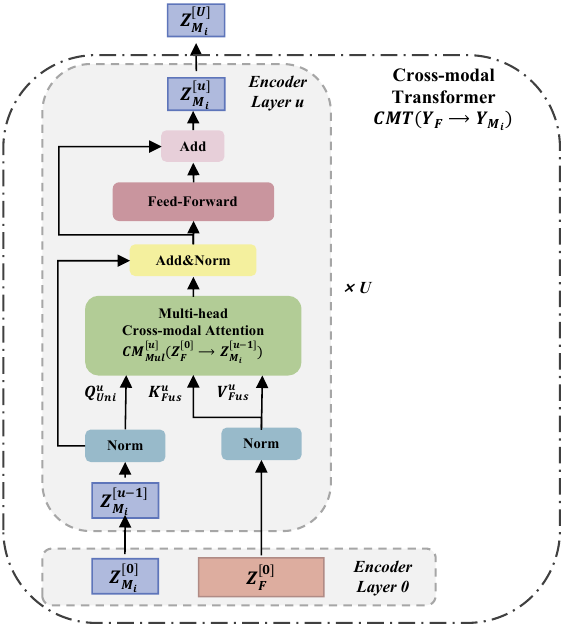}}
\caption{Architectural description of cross-modal attention and cross-modal transformer network.}
\label{fig:c}
\end{figure}

The purpose of our cross-modal attention is to learn an attention score between a target low-level unimodal feature sequence $Y_{M_i}\in \mathbb{R}^{L_{M_i},D}$ and the low-level multi-modal fusion representation $Y_{F} \in \mathbb{R}^{{L_{F}},D}$, which guides the adaption and reinforcement for the target unimodal features using other source unimodal features embedded in the fusion representation. We formulate unimodal Query $Q_{Uni}$, fusion Key $K_{Fus}$ and Value $V_{Fus}$ as: 
\begin{equation}
\label{Cross_qkv}
\begin{aligned}
Q_{Uni} &= Y_{M_i} \cdot W_{Q^{Uni}}\\
K_{Fus} &= Y_{F} \cdot W_{K^{Fus}}\\
V_{Fus} &= Y_{F} \cdot W_{V^{Fus}}
\end{aligned}
\end{equation}

\noindent where $W_{Q^{Uni}} \in \mathbb{R}^{D, D_Q}$, $W_{K^{Fus}} \in \mathbb{R}^{D, D_K}$ and $W_{V^{Fus}} \in \mathbb{R}^{D, D_V}$ are learnable weights.

As depicted in Figure \ref{fig:cm}, similar to the self-attention process described in \cite{vaswani2017attention}, the latent adaption and reinforcement from the fusion representation to the target unimodal feature, i.e., the learned cross-modal attention $\dot Y^{head_j}_{M_i} \in \mathbb{R}^{L_{M_i},D_V}$, in the $j^{th}$ head cross-modal attention can be defined as:
\begin{equation}
\label{eq_ca}
\begin{split}
\dot Y^{head_j}_{M_i} &= CM(Y_{F} \rightarrow Y_{M_i})\\
&= {Attention (Q^j_{Uni}, K^j_{Fus}, V^j_{Fus})}\\
&= \operatorname{softmax}\left(\frac{{Q^j_{Uni}} \cdot {K^j_{Fus}}^\top}{\sqrt{D_{K}}}\right)\cdot{V^j_{Fus}}
\end{split}
\end{equation}
\noindent where the $\operatorname{softmax}(.) \in \mathbb{R}^{L_{M_i},L_{F}}$  presents the scaled cross-modal attention score matrix between the fusion representation and the target unimodal feature.

We define the $\dot Y^{head_j}_{M_i}$ in Eq. (\ref{eq_ca}) as the single-head cross-modal attention. Accordingly, the multi-head cross-modal attention between the $i^{th}$ target modality and the fusion representation can be formulated as:
\begin{equation}
\label{mul_cm}
\begin{split}
{\dot Y^{Mul}_{M_i}} &= CM_{Mul} (Y_{F} \rightarrow Y_{M_i})\\
&=\operatorname{Concat}\left(\dot{Y}^{head_1}_{M_i},\ldots, \dot{Y}^{head_m}_{M_i}\right) 
\end{split}
\end{equation}
\noindent where $\dot Y^{Mul}_{M_i} \in \mathbb{R}^{L_{M_i},mD_V}$ and $m$ is the head number.

\subsubsection{Cross-modal Transformer}

Based on the structure of the self-attention transformer network in \cite{vaswani2017attention}, we build up the cross-modal transformer using the above-defined multi-head cross-modal attention. As shown in Figure \ref{fig:cmt}, the cross-modal transformer with the input of the $i^{th}$ target unimodal feature sequence and the fusion representation, $CMT_{Y_{F} \rightarrow Y_{M_{i}}}$, is composed of multiple identical cross-modal attention encoder layers, each of which contains a multi-head cross-modal attention block and a position-wise feed-forward network with residual connection and layer normalization. Generally, a cross-modal transformer network with $U$ cross-modal attention encoder layers calculates the reinforced unimodal feature sequence $Z_{M_i}\in \mathbb{R}^{L_{M_i},D}$ feed-forwardly as:
\begin{equation}
\label{eq_cmt}
\begin{split}
&Z_{F}^{[0]} = Y_{F}\\
&Z_{M_i}^{[0]} = Y_{M_i} \\
&\hat{Z}_{M_i}^{[u]} =  CM^{[u]}_{Mul}(\mathbb{L}(Z_{F}^{[0]}) \rightarrow \mathbb{L}(Z_{M_i}^{[u-1]}) )\\
&\dot{Z}_{M_i}^{[u]} = \hat{Z}_{M_i}^{[u]} + \mathbb{L}(Z^{[u-1]}_{M_i})\\
&{Z}_{M_i}^{[u]} = \mathbb{F}_{\theta}(\mathbb{L}(\dot{Z}_{M_i}^{[u]})) + \mathbb{L}(\dot{Z}_{M_i}^{[u]})
\end{split}
\end{equation}

\noindent where $u\in[1,U]$ presents the $u^{th}$ cross-modal attention encoder layer, $\mathbb{L} (.)$ and $\mathbb{F}_{\theta}(.)$ denote the layer normalization operation and position-wise feed-forward network with a parameter set $\theta$ respectively.

In each cross-modal transformer, the target unimodal features are continuously encoded and enhanced with external information from other source unimodal features embedded in the fusion representation. Specifically, the low-level unimodal features of source modalities from the low-level fusion representation are converted to different pairs of the fusion Keys and Values in Eq. (\ref{Cross_qkv}) to compute the multi-head cross-modal attention of the target modality in Eq. (\ref{mul_cm}). Following the process described in Eq. (\ref{eq_cmt}), each target modality is merged with other source modalities by a position-wise feed-forward process from each cross-modal transformer.

\subsection{Self-attention Module}
\label{SM}
The outputs of the cross-modal attention module, to wit the reinforced unimodal features $Z_{M_i}^{[U]}\in \mathbb{R}^{L_{M_i},D}$ of all modalities, are concatenated together as a sequence, i.e., the mid-level fusion representation ${Z}_{\{ {M_1}, \ldots,{M_n} \}} \in \mathbb{R}^{L_{{M_1}, \ldots,{M_n}},D}$, which is transported to the self-attention transformer \cite{vaswani2017attention} described in Section \ref{TN} to produce a weighted high-level fusion representation $Z_{F} \in \mathbb{R}^{L_{F},D}$ with meaningful contextual information highlighted. Specifically, the self-attention transformer assigns adaptive self-attention scores to different positions in the input sequence, i.e., each minimum unit of the reinforced unimodal features of different modalities, by integrating the global contextual information of the whole sequence. As a result, reinforced unimodal features in the sequence are given adaptive weights to generate the high-level global representation, where important 
contextual information for the recognition of the human state, e.g., sensitive and efficient features under a certain task scenario, is highlighted, while unimportant one, e.g., disrupted and insensitive features, is neglected. Such an adaptive self-attention process further reduces the potential feature noncommensurability across modalities and improves the efficiency of the model.

\subsection{Model Training}
\label{MT}
At the final step, the output of the self-attention module, namely, the high-level global feature $Z_{F} \in \mathbb{R}^{L_{F},D}$, is passed through two linear layers with a residual connection operation and a softmax nonlinear activation function that calculates prediction probabilities $P$ as:
\begin{equation}
\begin{split}
\hat{Z} &= Z_F + \zeta_\rho(Z_F)\\  
P &= \operatorname{softmax}(\zeta_\beta(\hat{Z}))
\label{p}
\end{split}    
\end{equation}

\noindent where $\zeta_\rho$ and $\zeta_\beta$ present two linear layers with parameter sets $\rho$ and $\beta$, and $P$ contains the prediction probability for each class.

To reduce the class imbalance resulting from biased data distribution and varied recognition difficulty, we adopt the multi-class focal loss function \cite{lin2017focal} for training, which is formulated as:
\begin{equation}
{FL}(y, p) = - \sum_{c=1}^{C} \alpha_c y_c (1 - p_c)^\gamma \log(p_c)
\label{eq_loss}
\end{equation}
\noindent where $C$ is the total number of classes, $y_c$ and $p_c$ correspond to the true label and predicted probability for class $c$ respectively, $\alpha_c$ and $\gamma$ present the balancing parameter that controls the trade-off between the positive and negative samples within class $c$ and a focusing parameter that down-weights the contribution of well-classified samples respectively.

The overall procedures of \textit{Husformer} model training are summarized in Algorithm \ref{alg:1}. 

\begin{algorithm}[htb]  
	\caption{Procedures of \textit{Husformer} Training }  		\label{alg:1}  
	\begin{algorithmic}[1]  
		\State Given multi-modal data series $X_{\{M_{1}, \ldots, M_n}\}$ of $n$ modalities  and true classification labels $y$
		\State Given training steps $T$
		\State Initialize convolution kernel sizes $k_{\{M_{1}, \ldots, M_n\}}$, cross-modal attention weights $W_{Q_{Uni}}, W_{K_{Fus}}, W_{V_{Fus}}$, self-attention weights $W_{Q}, W_{K}, W_{V}$ and other model parameters
		\State $t \leftarrow 0$
		\State \textcolor[RGB]{102,153,153}{// Convolutions and Positional Encoding}
		\State Compute low-level unimodal feature sequences $Y_{\{M_{1}, \ldots, M_n}\}$ by (\ref{eqconv})-(\ref{eq_pe})
		\State Compute low-level fusion representation $Y_{F}$ by (\ref{eq_fus})
		\While{$t<T$}
		\State \textcolor[RGB]{102,153,153}{// Cross-modal Attention Module}
		\State Compute reinforced unimodal feature sequences $Z_{\{M_{1}, \ldots, M_n}\}$ by (\ref{eq_cmt})
		\State \textcolor[RGB]{102,153,153}{// Self-attention Module}
		\State Compute high-level fusion representation $Z_F$ 
		\State \textcolor[RGB]{102,153,153}{// Linear Layers}
		\State Compute prediction probability $P$ by  (\ref{p})
		\State \textcolor[RGB]{102,153,153}{// Model Optimization}
		\State Compute loss $FL(y, p)$ by (\ref{eq_loss})
		\State Update model parameters using back propagation
		\State $t \leftarrow t+1$
        \EndWhile

	\end{algorithmic}  
\end{algorithm}

\section{Experiment and Results}
\label{experiment}
In this section, we describe the experiments we conducted on four public multi-modal datasets for human affective and cognitive state recognition. These datasets are widely used in the field, and we compared our \textit{Husformer} with four state-of-the-art baselines of multi-modal fusion-based human state recognition. For each dataset, the performance of using a single modality was also reported to investigate if our proposed multi-modal fusion-based \textit{Husformer} can outperform recognition methods using a single modality. Additionally, we conducted an ablation study to evaluate the benefits of each module in \textit{Husformer}.

\subsection{Datasets}
We selected two multi-modal affective datasets: DEAP \cite{koelstra2011deap} and WESAD \cite{schmidt2018introducing}, and two multi-modal cognitive load datasets: MOCAS  \cite{jo2022mocas} and CogLoad \cite{gjoreski2020datasets} for experiments.

The DEAP dataset contains multiple physiological signals, including EEG, EMG, EOG, and GSR, collected from 32 participants watching 40 different music video clips that elicited different emotional states. After each video clip, participants were requested to report their affective state levels in terms of arousal, valence, liking, and dominance from 1 to 9 using the Self-Assessment Manikin (SAM). In our experiment, we utilized two versions of the DEAP dataset: the downloaded \textit{Data-original.zip}, which contained collected raw multi-modal features, was regarded as the raw DEAP dataset, while the downloaded \textit{Data-preprocessed.zip}, which crafted features with several procedures\footnote{\url{https://www.eecs.qmul.ac.uk/mmv/datasets/deap/readme.html}},  was regarded as the preprocessed DEAP dataset. Valence and arousal were selected as the evaluation criteria of human emotion, where we mapped the scales (1-9) into three levels: “negative” or “passive"(1-3); “neutral" (4-6); and “positive" or “active" (7-9).
\begin{table}[t]
\centering
\caption{Description of utilized modalities in the DEAP datasets. Frequency: time sampling rate; Channels: number of channels; and Array Shape: number of data rows$\times$channels$\times$frequency. }
\label{tab:deap}
\begin{tabular}{cccc}
\hline
\multicolumn{4}{c}{\textbf{Raw DEAP Dataset}}      \\ 
Modality & Frequency & Channels & Array Shape        \\ \hline\hline
EEG      & 512        & 32             & {}69535$\times$32$\times$512{}    \\
EMG      & 512        & 4             & {}69535$\times$4$\times$512{}    \\
EOG      & 512      & 4         & {}69535$\times$4$\times$512{}    \\
GSR     & 512       & 1            & {}69535$\times$1$\times$512{}    \\ \hline \hline
\multicolumn{4}{c}{\textbf{Preprocessed DEAP Dataset}}        \\
Modality & Frequency & Channels & Array Shape         \\ \hline \hline
EEG      & 128       & 32             & {}80640$\times$32$\times$128{} \\
EMG      & 128       & 2              & {}80640$\times$2$\times$128{}  \\
EOG      & 128       & 2              & {}80640$\times$2$\times$128{}  \\
GSR       & 128       & 1              & {}80640$\times$1$\times$128{}  \\ \hline
\end{tabular}
\end{table}

\begin{table}[t]

\centering
\caption{Description of utilized modalities in the WESAD dataset}
\label{tab:wesad}
\begin{tabular}{cccc}
\hline
\multicolumn{4}{c}{\textbf{WESAD Dataset}}    \\
Modality & Frequency & Channels & Array Shape        \\ \hline \hline
GSR (chest)  & 700       & 1              & {}27287$\times$1$\times$700{} \\
BVP (wrist)  & 64        & 1              & {}27287$\times$1$\times$64{}  \\ 
EMG (chest)  & 700       & 1              & {}27287$\times$1$\times$700{} \\
ECG (chest)  & 700       & 1              & {}27287$\times$1$\times$700{} \\
RESP (chest) & 700       & 1              & {}27287$\times$1$\times$700{} \\
GSR (wrist)  & 4         & 1              & {}27287$\times$1$\times$4{}   \\\hline 
\end{tabular}
\end{table}
The WESAD dataset contains physiological data, consisting of GSR, BVP, EMG, ECG and respiration (RESP), collected by one chest-worn and one wrist-worn wearable sensor from 15 participants who conducted different tasks that aimed to elicit different emotional states. Specifically, participants were asked to close their eyes for seven minutes to stimulate the neutral state. The stress state was elicited by the Trier Social Stress Test (TSST) \cite{kirschbaum1993trier}, where participants delivered a five-minute speech on their personal traits to three-person panels. Moreover, for the amusement state, participants were required to watch funny videos for 392 seconds. After each task, participants reported subjective emotional states using SAM and other self-report questionnaires, and three kinds of labels, i.e., neutral \textit{vs.} stress \textit{vs.} amusement, were provided.

The MOCAS dataset contains physiological data, including 5-channel EEGs, EEG band powers (or EEG\_POW, including theta, low and high beta, alpha, and gamma bands of 5-channel EEGs), BVP, GSR and HR, and behavioral data, including Eye Aspect Ratio (EAR) and Action units (AUs), from 21 participants conducting Closed-Circuit Television (CCTV) monitoring tasks that aimed to elicit different levels of cognitive load. After each task, participants were required to report subjective cognitive load via NASA-TLX. Based on the weighted NASA-TLX scores, three categories of annotations, i.e., low \textit{vs.} medium \textit{vs.} high, of cognitive load were given. Apart from the raw multi-modal physiological and behavioral features collected from two off-the-shelf wearable sensors: Empatica E4 and Emotiv Insight, and a webcam, the MOCAS also contains the data prepossessed by NeuroKit2 \cite{makowski2021neurokit2} and other methods \cite{jo2022mocas}. In this experiment, we utilized both raw and preprocess MOCAS dataset.

\begin{table}[t]
\centering
\caption{Description of utilized modalities in the MOCAS datasets}
\label{tab:mocas}
\begin{tabular}{cccc}
\hline
\multicolumn{4}{c}{\textbf{Raw MOCAS Dataset}}           \\
Modality & Frequency & Channels & Array Shape         \\ \hline \hline
EEG         & 128   & 6     & {}215341$\times$5$\times$128{}   \\
EEG\_POW    & 8     & 25    & {}215341$\times$25$\times$8{}  \\
BVP         & 128   & 1     & {}215341$\times$1$\times$128{} \\
GSR         & 6     & 1     & {}215341$\times$1$\times$6{}  \\
EAR         & 1    & 1     & {}215341$\times$1$\times$1{}  \\ \hline \hline
\multicolumn{4}{c}{\textbf{Preprocessed MOCAS Dataset}}  \\
Modality & Frequency & Channels & Array Shape         \\ \hline\hline
EEG         & 128   & 6     & {}215341$\times$5$\times$128{}  \\
EEG\_POW    & 8     & 25    & {}215341$\times$25$\times$8{}   \\
BVP         & 128   & 1     & {}215341$\times$1$\times$128{} \\
GSR         & 6     & 1     & {}215341$\times$1$\times$6{}  \\
EAR         & 1    & 1     & {}215341$\times$1$\times$1{}  \\ \hline
\end{tabular}
\end{table}
\begin{table}[t]
\centering
\caption{Description of utilized modalities in the CogLoad dataset}
\label{tab:CogLoad}
\begin{tabular}{cccc}
\hline
\multicolumn{4}{c}{\textbf{CogLoad Dataset}}   \\
Modality & Frequency & Channels & Array Shape      \\ \hline\hline
HR       & 1         & 1              & {}89225$\times$1$\times$1{} \\
IBI      & 1         & 1              & {}89225$\times$1$\times$1{} \\
GSR      & 1         & 1              & {}89225$\times$1$\times$1{} \\
SKT      & 1         & 1              & {}89225$\times$1$\times$1{} \\
ACC      & 1         & 2              & {}89225$\times$2$\times$1{} \\ \hline
\end{tabular}
\end{table}

The CogLoad dataset contains physiological signals, including HR, IBI, GSR, SKT, and motion data (ACC) collected from 23 participants through a Microsoft band. The participants conducted six dual tasks including primary and secondary cognitive-load tasks that were expected to stimulate target levels of cognitive load. The primary task was randomly selected from six psycho-physiological tests proposed in \cite{haapalainen2010psycho}. The secondary task was to click on the appearing target on screen while conducting the primary task.
After each task, participants were asked to report subjective cognitive load based on the TLX mental demand of NASA-TLX questionnaire, and the data collected in the baseline (rest) section was labelled as -1. In our experiment, we mapped the subjective scales into three classes of labels as low (-1-3), medium (4-6) and high (7-9). 

In our experiments, to simulate realistic application scenarios, the input sample of each modality is a 2-D feature matrix extracted from a 1-second segment with the dimension of ${L_{M_i} \times D_{M_i}}$, i.e., the channel number plus the sampling frequency of the modality. The details of the modalities used in the aforementioned datasets are described in Table \ref{tab:deap}-\ref{tab:CogLoad}.

\subsection{Baselines}
We selected four state-of-the-art baselines of multi-modal fusion-based human state recognition as the comparisons with our proposed \textit{Husformer}:

\begin{itemize}
    \item \textit{EF-SVM}: Support Vector Machines (SVMs) with early fusion \cite{hogervorst2014combining,zhang2017feature,zhou2020multimodal}. This is a popular benchmark model for human state recognition, which concatenates different modalities together at the sensor or feature level and builds a SVM as the classifier for the fused representation. 
    
    \item \textit{LF-SVM}: SVMs with late fusion \cite{putze2010multimodal,zhang2017feature,zhou2020multimodal}. This model is also a strong benchmark to predict human state, which fuses different modalities at the decision level. Each modality is processed with a SVM classifier to make individual predictions, which are combined together through a voting scheme. In our experiment, we select Dempster-Shafer Theory (DST) voting \cite{zhou2018early,shafer1992dempster}, which is reported as the best performing voting scheme by \cite{zhou2020multimodal}, as the late fusion process.

    \item \textit{EmotionMeter} \cite{zheng2018emotionmeter}. This is a multi-modal deep learning model for human emotion prediction. Multiple individual RBMs are built to process features of each modality, where the hidden layers of those individual RBMs are concatenated together to learn the shared representations across different modalities. Then a linear SVM is adopted for classification using the learned shared representations.

    \item \textit{MMResLSTM}: Multimodal Residual Long Short-Term Memory Neural Network \cite{ma2019emotion}. This is another state-of-the-art multi-modal deep learning model for human emotion recognition. Multiple individual LSTM blocks are constructed for each modality, where each LSTM layer of these individual LSTM blocks shares the same weights to learn the temporal correlations across different modalities. Finally, the outputs of all LSTM blocks are concatenated and passed through a dense layer to make predictions. Layer normalization and residual connection are also applied to accelerate training process.
\end{itemize}

Moreover, to comprehensively evaluate if our \textit{Husformer} could improve the performance compared with using single modality, we also implemented Long Short-Term Memory Nerual Network (LSTM) \cite{graves2012long}, Graph Neural Network (GNN) \cite{scarselli2008graph} and Transformer \cite{vaswani2017attention}, which were broadly utilized for single-modal-based human state recognition \cite{song2018eeg,zhong2020eeg,sun2021eeg}, to test each single modality in each dataset, and the best performing results among these three models were reported.

\subsection{Ablation Study}
To evaluate the benefits of each module in our \textit{Husformer}, we also built three ablation models for the ablation study:

\begin{itemize}
    \item \textit{HusFuse}: Deleting the cross-modal attention module in the \textit{Husformer}, and fusing the low-level features of all modalities directly to the self-attention module.
    \item \textit{HusLSTM}: Replacing the self-attention transformer in the \textit{Husformer} with an LSTM layer.
    \item \textit{HusPair}: Replacing the cross-modal attention module in the \textit{Husformer} with the directional pairwise cross-modal attention widely adopted in multi-modal natural language processing and computer vision areas \cite{hu2021unit,tsai2019multimodal}. However, \textit{Husformer} differs from \textit{HusPair} in that it focuses on the cross-modal attention between each individual modality and the multi-modal fusion signal, rather than between one single modality and another. This allows the model to consider the coordination of more than a pair of modalities at the same time, reducingthe potential information redundancy caused by parallel pairwise fusion. 
\end{itemize}

\subsection{Evaluation and Metrics}  
For each dataset, we randomly shuffled all data and conducted the $K$-folder cross validation ($K=10$) \cite{russell2010artificial,gareth2013introduction}. We reported the average multi-class accuracy ($Acc$) and multi-class average F1-score ($F1$) \cite{bishop2006pattern} with standard deviations for each model in each experiment. Furthermore, samples from the same trials, such as a film clip in the DEAP dataset or a monitoring task in the MOCAS dataset, were exclusively included in either the training or test sets.

Assuming the number of classes in a multi-class classification task is $n$, we can define $Acc$ as:
\begin{equation}
\text {$Acc$}=\frac{\sum_{i=1}^{n} accuracy_{i}}{n}
\end{equation}
\noindent where $accuracy_{i}$ denotes the binary accuracy of the $i^{th}$ class as:
\begin{equation}
{accuracy_{i}}=\frac{{TP_{i}}+{TN_{i}}}{lenth_{i}}
\end{equation}
\noindent where ${TP_{i}}$, ${TN_{i}}$ and $lenth_{i}$ present true positive samples, true negative samples, and the number of total samples in the $i^{th}$ class, respectively.

The $F1$ can be formulated as:
\begin{equation}
\text {$F1$}=\frac{\sum_{i=1}^{n} f1_{i} }{n}
\end{equation}
\noindent where $f1_{i}$ denotes the binary F1 score of the $i^{th}$ class as:
\begin{equation}
{f1_{i}}=\frac{2 \times precision_{i}\times recall_{i}}{precision_{i}+recall_{i}}
\end{equation}

\begin{equation}
{precision_{i}}=\frac{TP_{i}}{{TP_{i}}+{FP_{i}}}
\end{equation}

\begin{equation}
{recall_{i}}=\frac{TP_{i}}{{TP_{i}}+{FN_{i}}}
\end{equation}

\noindent where $precision_{i}$ and $recall_{i}$ present the binary precision and recall separately, ${FP_{i}}$ and ${FN_{i}}$ denote false positive and false negative samples in the $i^{th}$ class respectively. 

We also conducted two-sample independent t-tests to compare the classification results of different models, and the significance is asserted when $p < 0.01$. 

\subsection{Implementation Details}
All training and experiments were conducted on a NVIDIA Tesla V100 GPU. We trained all baseline networks by following the implementation procedures described in their respective original papers. Also, for the SVM classifier, we followed the approach descrbied in \cite{zhou2020multimodal} to select the Radial Basis Function (RBF) kernel and optimize the values of \textit{C} and $\gamma$. Note that \textit{EF-SVM} cannot be directly applied to unaligned datasets, which means datasets that contain multiple modalities with different time sampling rates, since the concatenation operation is mathematically impossible due to different feature dimensions. Therefore, we added multiple one-dimensional convolution sub-networks with the same structures and parameters as those in the \textit{Husformer} before the \textit{EF-SVM} to extract low-level unimodal features with the same dimension for the concatenation operation on unaligned datasets, i.e., WESAD and MOCAS datasets. Furthermore, to ensure a fair comparison, we kept the hyper-parameters of ablation models the same as those in the \textit{Husformer} during the experiments. The hyper-parameters of the \textit{Husformer} used in each experiment can be found in Appendix \ref{appendix:A}.


\subsection{Quantitative Measurements}
\begin{table*}[htb]
\centering
\caption{Performance of different models on raw DEAP and prepossessed DEAP datasets in terms of average multi-class average accuracy ($Acc$) and multi-class average F1-score ($F1$) with stand deviations. Results of other models that are within $5\%$ of \textit{Husformer}'s performance on $Acc$ or $F1$ are highlighted. $^h$: higher values indicate better performance.}
    \label{tab:DEAP}
\begin{tabular}{c||cccc||cccc}
\hline
{Dataset} & \multicolumn{4}{c||}{Raw DEAP}                                     & \multicolumn{4}{c}{Preprocessed DEAP}                            \\ \hline \hline
{Criteria}    & \multicolumn{2}{c}{Valence}     & \multicolumn{2}{c||}{Arousal}    & \multicolumn{2}{c}{Valence}     & \multicolumn{2}{c}{Arousal}    \\
{Metric}      & $Acc(\%)^h$           & $F1(\%)^h$           & $Acc(\%)^h$           & $F1(\%)^h$             & $Acc(\%)^h$          & $F1(\%)^h$            & $Acc(\%)^h$          & $F1(\%)^h$             \\ \hline \hline

EF-SVM                         & 43.95$\pm$2.17 & 47.36$\pm$2.53 & 46.02$\pm$2.10 & 48.69$\pm$2.16 & 70.68$\pm$6.30 & 72.40$\pm$6.52 & 71.04$\pm$5.97 & 71.18$\pm$6.14 \\
LF-SVM                         & 45.09$\pm$4.82 & 49.90$\pm$5.18 & 48.18$\pm$4.01 & 51.40$\pm$3.96 & 67.59$\pm$5.60 & 69.37$\pm$5.77 & 70.24$\pm$4.95 & 71.11$\pm$5.09 \\
EmotionMeter                   & 61.71$\pm$3.45 & 62.00$\pm$3.39 & 62.08$\pm$3.16 & 62.18$\pm$3.08 & 85.26$\pm$2.52 & 79.59$\pm$2.70 & 80.02$\pm$3.32 & 80.18$\pm$3.25 \\
MMResLSTM                      & 65.68$\pm$2.13 & 66.39$\pm$2.05 & 66.31$\pm$1.78 & 66.39$\pm$1.86 & \textbf{86.78$\pm$2.56} & \textbf{87.03$\pm$2.55} & \textbf{86.55$\pm$1.78} & \textbf{87.13$\pm$2.25} \\   
\hline \hline
HusFuse                        & 67.45$\pm$3.23 & 68.64$\pm$3.16 & 67.85$\pm$2.67 & 68.08$\pm$2.53 & 80.48$\pm$1.58 & 80.77$\pm$1.71 & 81.26$\pm$1.38 & 81.42$\pm$1.48 \\
HusLSTM                        & 72.66$\pm$2.34 & 73.09$\pm$2.37 & 71.03$\pm$1.90 & 71.40$\pm$1.93 & 83.41$\pm$1.90 & 84.15$\pm$2.09 & 84.61$\pm$1.58 & 84.73$\pm$1.47 \\ 
HusPair                        & \textbf{77.14$\pm$2.40} & \textbf{76.71$\pm$2.18} & \textbf{77.55$\pm$2.22} & \textbf{77.05$\pm$2.09} & \textbf{89.42$\pm$3.33} & \textbf{89.26$\pm$2.99} & \textbf{90.31$\pm$2.99} & \textbf{90.15$\pm$3.06} \\
\hline\hline
HusFormer                      & \textbf{79.64$\pm$1.52} & \textbf{79.87$\pm$1.54} & \textbf{79.94$\pm$2.18} & \textbf{80.44$\pm$2.25} & \textbf{90.67$\pm$2.20} & \textbf{90.74$\pm$2.29} & \textbf{91.33$\pm$1.59} & \textbf{91.35$\pm$1.67} \\ \hline
\end{tabular}
\end{table*}

\begin{table*}[htb]
\centering
\caption{Performance of different models on WESAD, raw MOCAS, preprocessed MOCAS, and CogLoad datasets in terms of average multi-class average accuracy ($Acc$) and multi-class average F1-score ($F1$) with stand deviations. Results of other models that are within $5\%$ of the \textit{Husformer}'s performance on $Acc$ or $F1$ are highlighted. $^h$: higher values indicate better performance.}
    \label{tab:others}
\begin{tabular}{c||cc||cc||cc||cc}
\hline
Dataset      & \multicolumn{2}{c||}{WESAD}                & \multicolumn{2}{c||}{Raw MOCAS}            & \multicolumn{2}{c||}{Preprocessed MOCAS}   & \multicolumn{2}{c}{CogLoad}    \\ \hline\hline
Metric           & $Acc(\%)^h$           & \multicolumn{1}{c||}{$F1(\%)^h$} & $Acc(\%)^h$            & \multicolumn{1}{c||}{$F1(\%)^h$} & $Acc(\%)^h$            & \multicolumn{1}{c||}{$F1(\%)^h$} & $Acc(\%)^h$           & $F1(\%)^h$           \\ \hline\hline

EF-SVM           & 42.46$\pm$4.34 & 44.39$\pm$4.08           & 51.48$\pm$4.39 & 51.63$\pm$5.00           & 62.73$\pm$4.91 & 61.87$\pm$4.19           & 41.67$\pm$3.80 & 47.52$\pm$3.14 \\
LF-SVM           & 44.98$\pm$2.48 & 47.51$\pm$3.00           & 48.74$\pm$3.40 & 48.85$\pm$3.42           & 59.80$\pm$5.16 & 60.68$\pm$5.07           & 38.98$\pm$2.71 & 45.87$\pm$2.12 \\
EmotionMeter     & 63.01$\pm$1.41 & 63.21$\pm$1.34           & 71.15$\pm$3.48 & 70.98$\pm$3.39           & 78.80$\pm$2.54 & 79.94$\pm$2.61           & 59.57$\pm$1.42 & 62.99$\pm$1.30 \\
MMResLSTM        & 65.76$\pm$1.12 & 66.32$\pm$1.24           & 75.33$\pm$2.41 & 75.44$\pm$2.21           & 82.81$\pm$1.34 & 83.25$\pm$1.40           & 61.44$\pm$1.67 & 63.39$\pm$1.71 \\ \hline\hline
HusFuse          & 68.77$\pm$1.56 & 68.48$\pm$1.31           & 70.65$\pm$2.36 & 71.22$\pm$2.39           & 78.00$\pm$2.10 & 78.81$\pm$1.86           & 58.49$\pm$0.68 & 57.65$\pm$0.83 \\
HusLSTM          & 70.64$\pm$1.21 & 71.00$\pm$1.28           & 78.98$\pm$2.72 & 79.28$\pm$2.61           & 82.40$\pm$1.80 & 82.54$\pm$1.78           & 67.09$\pm$1.06 & 66.60$\pm$1.02 \\ 
HusPair          & 73.57$\pm$1.72 & 73.77$\pm$2.13           & 82.12$\pm$1.83 & 82.46$\pm$1.63           & \textbf{88.83$\pm$3.97} & \textbf{88.75$\pm$3.99}           & 65.11$\pm$3.07 & 66.55$\pm$3.25 \\

\hline\hline

HusFormer        & \textbf{78.68$\pm$2.05} & \textbf{79.51$\pm$2.28}           & \textbf{87.37$\pm$2.40} & \textbf{87.47$\pm$2.55}           & \textbf{90.09$\pm$2.25} & \textbf{90.17$\pm$2.17}           & \textbf{74.06$\pm$2.48} & \textbf{74.93$\pm$2.77} \\ \hline
\end{tabular}
\end{table*}

\subsubsection{Comparative Results with Baselines}
Table \ref{tab:DEAP} and \ref{tab:others} summarize the performance of our \textit{Husformer} when compared with the four multi-modal baselines of human state recognition in terms of average $Acc$ and $F1$ with stand deviations during the experiments. From the comparative results, we can observe that regarding the overall performance, our proposed \textit{Husformer} outperforms the other four state-of-the-art baselines in the recognition of both human affective state (DEAP and WESAD datasets) and cognitive load (MOCAS and CogLoad datasets), which demonstrates that the \textit{Husformer} can serve as a more effective backbone network for general human state prediction tasks. Also, such performance enhancements are more evident on datasets containing raw multi-modal features (raw DEAP and MOCAS, WESAD and CogLoad datasets), which shows that the \textit{Husformer} can learn from raw multi-modal features series more efficiently, and thus is more applicable to real-world task scenarios where extensive feature crafting and alignment is quite impractical and expensive. These performance improvements mainly result from following reasons: 

\begin{itemize}
    
    \item Experimental results show that our \textit{Husformer} significantly outperforms \textit{EF-SVM} and \textit{LF-SVM} regarding $Acc$ and $F1$ on each dataset. This is reasonable since these two methods generate fusion representations by a simple concatenation of multiple modalities at the feature or decision level, which ignores correlations across different modalities, and may be limited by the curse of dimensionality \cite{wu2014survey}, while the \textit{Husformer} fuses all unimodal features through a feed-forward process from multiple cross-modal attention transformers, where the complementary interactions among different modalities are considered sufficiently. Such neglect of cross-modal interactions even causes the performance of EF-SVM and LF-SVM to underperform compared to EmotionMeter and MMResLSTM by approximately 10\% in terms of accuracy ($Acc$) and F1 score ($F1$). These comparative results highlight the significance of awareness of correlations across multiple modalities for multi-modal human state recognition. Moreover, the SVM, which serves as the backbone model in both \textit{EF-SVM} and \textit{LF-SVM}, has been shown to have low efficiency when dealing with a large amount of input data \cite{hou2018research}. Additionally, it is quite sensitive to missing data, outliers, and noise \cite{wu2007robust}, making it heavily reliant on careful data cleaning and feature selection procedures. Therefore, while some studies demonstrate that \textit{EF-SVM} and \textit{LF-SVM} are sufficient to estimate human state with a high volume of data cleaning and careful feature engineering \cite{ma2019emotion,zhou2020multimodal}, it is reasonable to conduct that \textit{EF-SVM} and \textit{LF-SVM} may not perform well when dealing with relatively raw multi-modal features.

    \item The experimental results show that our proposed \textit{Husformer} significantly outperforms \textit{EmotionMeter} in terms of $Acc$ and $F1$ on all datasets. We argue that this is due to the fact that although \textit{EmotionMeter} considers cross-modal interactions by building shared hidden layers to learn shared representations, the constructed RBMs neglect time-related cross-modal interactions, which are critical as the temporal characteristic is a vital attribute of signals reflecting human state \cite{debie2019multimodal,ma2019emotion}. On the other hand, \textit{MMResLSTM} achieves better performance than \textit{EmotionMeter} and even performs competitively with our \textit{Husformer} on the preprocessed DEAP and preprocessed MOCAS datasets. This is because \textit{MMResLSTM} constructs LSTM layers that share the same weights for different modalities to learn shared representations, which can model the temporal cross-modal correlations effectively. However, on the other four datasets that involve minimal or no feature engineering, \textit{Husformer} consistently and significantly outperforms the \textit{MMResLSTM} in terms of both $Acc$ and $F1$. We argue that the reason lies in the fact that our proposed cross-modal attention mechanism considers cross-modal interactions by encouraging one modality to directly engage in the unimodal features of other modalities where strongly complementary representation information is presented to enhance itself. Such an attention-based fusion strategy can capture complicated and long-term complementary interactions across multiple modalities more efficiently, especially when dealing with raw multi-modal features involving less feature engineering.
    
    \item The performance improvements of our proposed \textit{Husformer} compared to the other four baselines become more pronounced when transitioning from datasets that involve feature engineering preprocessing to raw datasets. This can be attributed to the cross-modal attention and self-attention modules utilized in the \textit{Husformer}. Specifically, in addition to modeling the complicated complementary multi-modal interactions from raw multi-modal signals efficiently, the cross-modal transformers can neglect irrelevance, such as interrupted or insensitive low-level unimodal features, when calculating adaptive cross-modal attention. Moreover, the self-attention process further prioritizes the important information and diminishes the irrelevance and interruptions by assigning adaptive weights to each element in the sequence consisting of reinforced unimodal features of different modalities. Compared to the baselines, these two adaptive attention mechanisms, at the fusion process and high-level feature level, respectively, can help the model to distill effective representations of human state without being disturbed by potential feature noncommensurability and interruptions in the raw multi-modal features to a greater extent.
\end{itemize}

\subsubsection{Comparative Results with Single Modality}
\label{Single}

\begin{figure}[t]
\centering
{\includegraphics[width=0.95\linewidth]{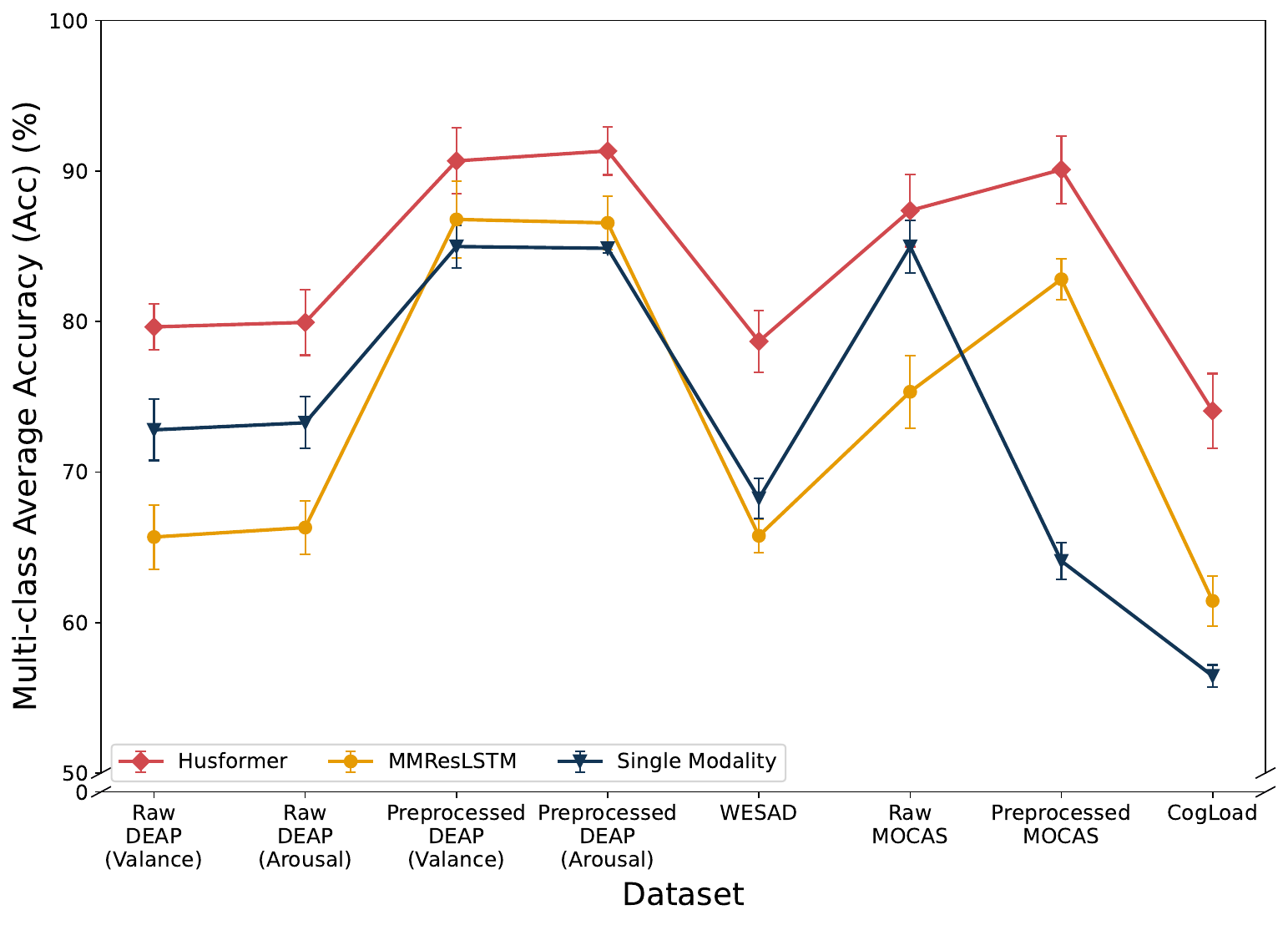}}
\caption{Performance comparison of our proposed \textit{Husformer} and the best-performing multi-modal fusion baseline, \textit{MMResLSTM}, with the highest recognition result achieved using a single modality on each dataset in terms of accuracy ($Acc$).}
\label{fig:single}
\end{figure}

Figure \ref{fig:single} presents a comparison of the performance of our proposed \textit{Husformer} and the best-performing multi-modal fusion baseline, \textit{MMResLSTM}, with the highest recognition result achieved using a single modality on each dataset, in terms of accuracy ($Acc$). Details of the classification outcomes for each individual modality in every dataset, with respect to both accuracy ($Acc$) and F1 score ($F1$), can be found in Appendix \ref{appendix:B}. As shown in Figure \ref{fig:single}, our proposed \textit{Husformer} significantly outperforms the best-performing single-modal-based recognition results on all four datasets in terms of $Acc$. We assume that these improvements result from the fact that our \textit{Husformer} can effectively leverage the aforementioned advantages of multi-modal fusion for human state recognition, such as combing metrics from different sources to unveil essential representation information that cannot be obtained from one single source \cite{lahat2015multimodal} and reducing the noise-to-signal ratio \cite{debie2019multimodal}. Additionally, we  observe that the \textit{MMResLSTM}, another multi-modal fusion-based model that achieves the best performance among all baselines, does not outperform the single EEG-related modality-based recognition with transformer networks on the raw DEAP and preprocessed MOCAS datasets. This demonstrates that the cross-modal attention and self-attention processes in our \textit{Husformer} enable it to take greater advantage of multi-modal fusion by modeling sufficient and long-term complementary cross-modal correlations and adaptively highlighting meaningful contextual representations.

\subsubsection{Results of the Ablation Study}
\label{ABS}
Table \ref{tab:DEAP} and \ref{tab:others} present the performance of the \textit{Husformer} in terms of $Acc$ and $F1$ on each dataset, compared to three ablation models, the \textit{HusFuse}, \textit{HusPair}, and \textit{HusLSTM}. From the results, we can observe the effectiveness of the cross-modal and self-attention module inside the \textit{Husformer} as follows:

\begin{table*}[htb]
\centering
\caption{The number of parameters and GPU memory usage during training of the \textit{Husformer} and \textit{Huspair} on each dataset. Note that both models had the same batch size during training. Para: the number of parameters; Mem: GPU memory usage.}
\label{para}
\resizebox{\linewidth}{!}{%
\begin{tabular}{c||cc||cc||cc||cc||cc||cc}
\hline
Dataset   & \multicolumn{2}{c||}{Raw DEAP} & \multicolumn{2}{c||}{Preprocessed DEAP} & \multicolumn{2}{c||}{WESAD} & \multicolumn{2}{c||}{Raw MOCAS} & \multicolumn{2}{c||}{Preprocessed MOCAS} & \multicolumn{2}{c}{Cogload} \\ \hline\hline
Metric    & Para        & Mem            & Para              & Mem                & Para        & Mem          & Para          & Mem            & Para               & Mem                & Para         & Mem          \\ \hline\hline
HusPair   & 2.90M      & 3253.24MB      & 2.92M           & 2908.66MB          & 3.90M     & 5416.90MB    & 6.21M       & 519.67MB       & 6.22M            & 531.80MB           & 3.12M      & 202.47MB      \\
Husformer & 0.63M       & 1999.56MB      & 0.66M           & 1773.59MB          & 0.71M      & 3084.35MB    & 0.74M       & 210.97MB       & 0.75M             & 220.53MB           & 0.72M       & 94.28MB       \\ \hline
\end{tabular}
}
\end{table*}

\begin{figure}[t]
    \centering
    \includegraphics[width=0.9\linewidth, height=0.28\textwidth]{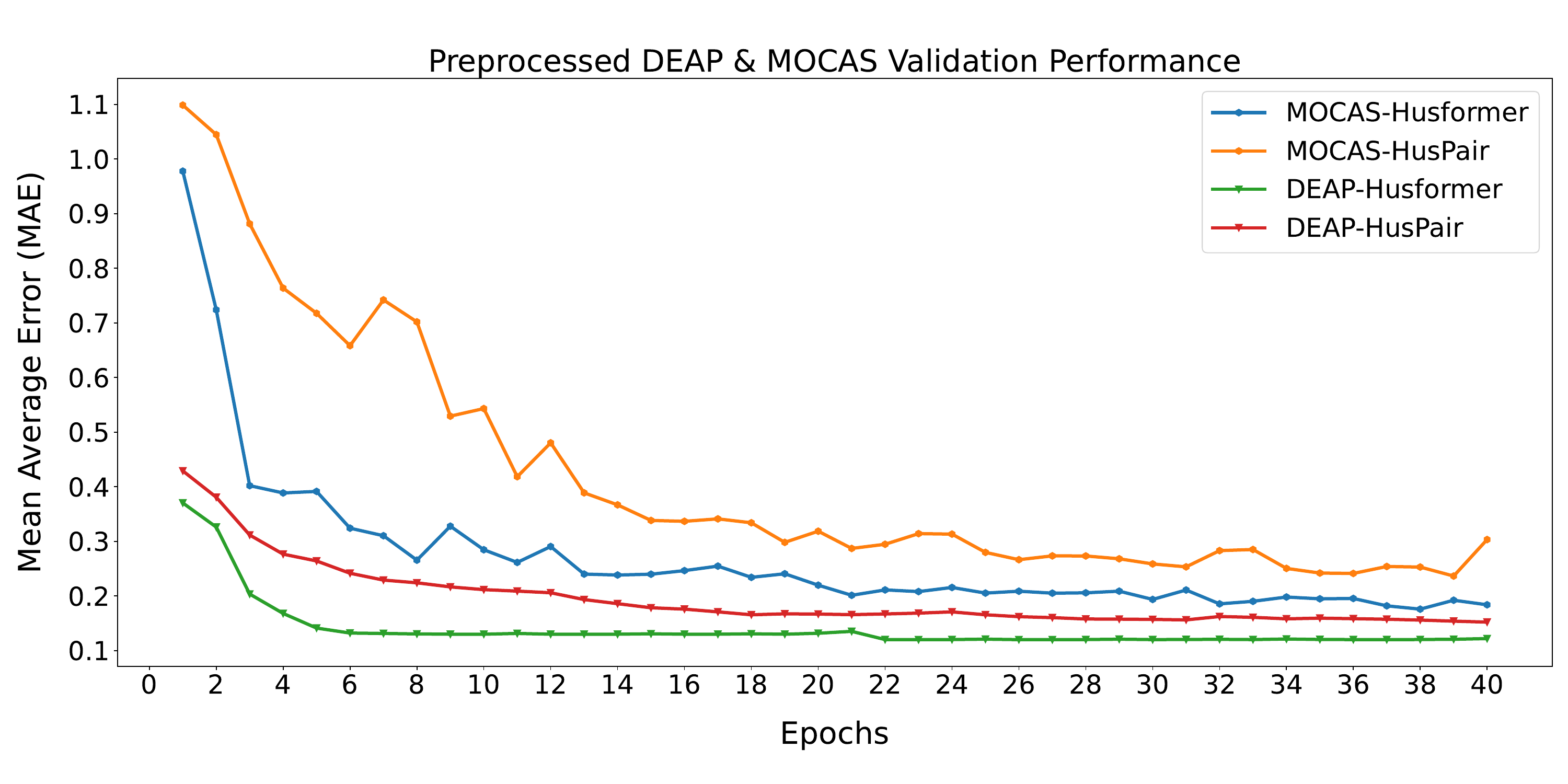}
    \caption{Learning curves of \textit{Husformer} when compared to \textit{HusPair} on the preprocessed DEAP and MOCAS datasets in terms of validation set convergence.}
    \label{fig:curve_MOCAS}
    \end{figure}

\textbf{Effectiveness of the cross-modal attention module}
\begin{itemize}
    \item Compared to the \textit{HusFuse}, which removes the cross-modal attention module from the \textit{Husformer}, the \textit{Husformer} achieves an absolute improvement in terms of $Acc$ and $F1$. This improvement can be attributed to the fact that the \textit{HusFuse} fuses low-level features from different modalities together using a simple concatenation operation, instead of utilizing a feed-forward process from cross-modal transformers. While the self-attention processes in the \textit{HusFuse} can be viewed as a way to consider cross-model ineractions, by relating entities of the sequence concatenated from low-level features of different modalities to contextual information to calculate the high-level fusion representation, it fails to provide direct complementary adaptions for features of one modality with those of other modalities during the fusion process. This result highlights the effectiveness of the proposed cross-modal attention-based fusion strategy  compared to simple concatenation with self-attention.

     \item Compared with the \textit{HusPair}, which replaces the cross-modal attention module in the \textit{Husformer} with the directional pair-wise cross-modal attention in \cite{tsai2019multimodal}, the \textit{Husformer} achieves an improvement in $Acc$ and $F1$. We argue that this is because the pairwise cross-modal attention in the \textit{HusPair} can only consider the complementary interactions between a pair of modalities at once and thus ignores the coordination among more than two modalities. In contrast, our proposed cross-modal attention computes the complementary interactions between the low-level unimodal features of one target modality and the low-level fusion representation embedded with unimodal features of the other source modalities. This allows the model to consider the coordination across all modalities at the same time, hence considering more long-term and comprehensive cross-modal interactions. Moreover, it has been shown that the pairwise fusion approach can produce redundant fusion information that may serve as additional noise rather than effectual multi-modal features \cite{bian2021attention,bhojanapalli2021leveraging}.

    \item We can also observe that \textit{HusPair} is quite comparable to the \textit{Husformer} on the DEAP datasets and preprocessed MOCAS dataset. However, as presented in Table \ref{para}, the parameter number of the \textit{HusPair} is about 4$\sim$8 times that of the \textit{Husformer}. This is because the number of the pairwise cross-modal attention transformers in the \textit{HusPair} increases exponentially with the increase in the number of modalities. Specifically, when applied to multi-modal fusion of $n$ modalities, the pairwise cross-modal attention requires $n^2-n$ cross-modal transformers, while ours only requires $n$ of them. For instance, on the WESAD dataset that contains six modalities, the \textit{HusPair} requires $30$ cross-modal transformers, while our \textit{Husformer} only requires $6$ of them. Such a high volume of parameters can result in slow convergence and high training difficulty. For example, on the preprocessed DEAP and the preprocessed MOCAS, where the \textit{HusPair} gets its most competitive results, we empirically observe that the \textit{Husformer} can converge faster to a lower loss of mean absolute error compared to the \textit{HusPair} during the training process (see Figure \ref{fig:curve_MOCAS}). Furthermore, we assume that the pairwise cross-modal attention in \textit{HusPair} with potential over-parameterization issues may also produce attention redundancy when the feature complexity cannot meet that of the attention, e.g., regarding the noise or artifacts as a part of the learned attention. For example, on the CogLoad dataset, where the feature length (channel number) and dimension (sampling frequency) are very shallow compared to those in other datasets, the performance of the \textit{HusPair} drops significantly, exceeded by the \textit{Husformer} by $10\%$ (see Table \ref{tab:DEAP} and \ref{tab:others}). Therefore, the high storage and computation costs (see Table \ref{para}), and the potential ineffectiveness on shallow modalities due to the large number of parameters limit the applicability of the \textit{HusPair} to general and practical scenarios of multi-modal human state recognition. On the other hand, our \textit{Husformer} achieves comparable and even significantly superior performance with much fewer parameters than the \textit{HusPair}. These results highlight the effectiveness of our proposed cross-modal attention compared to the directional pairwise attention approach. 
    \end{itemize}
    
\begin{figure*}[ht]
\centering
\includegraphics[width=1.8\columnwidth]{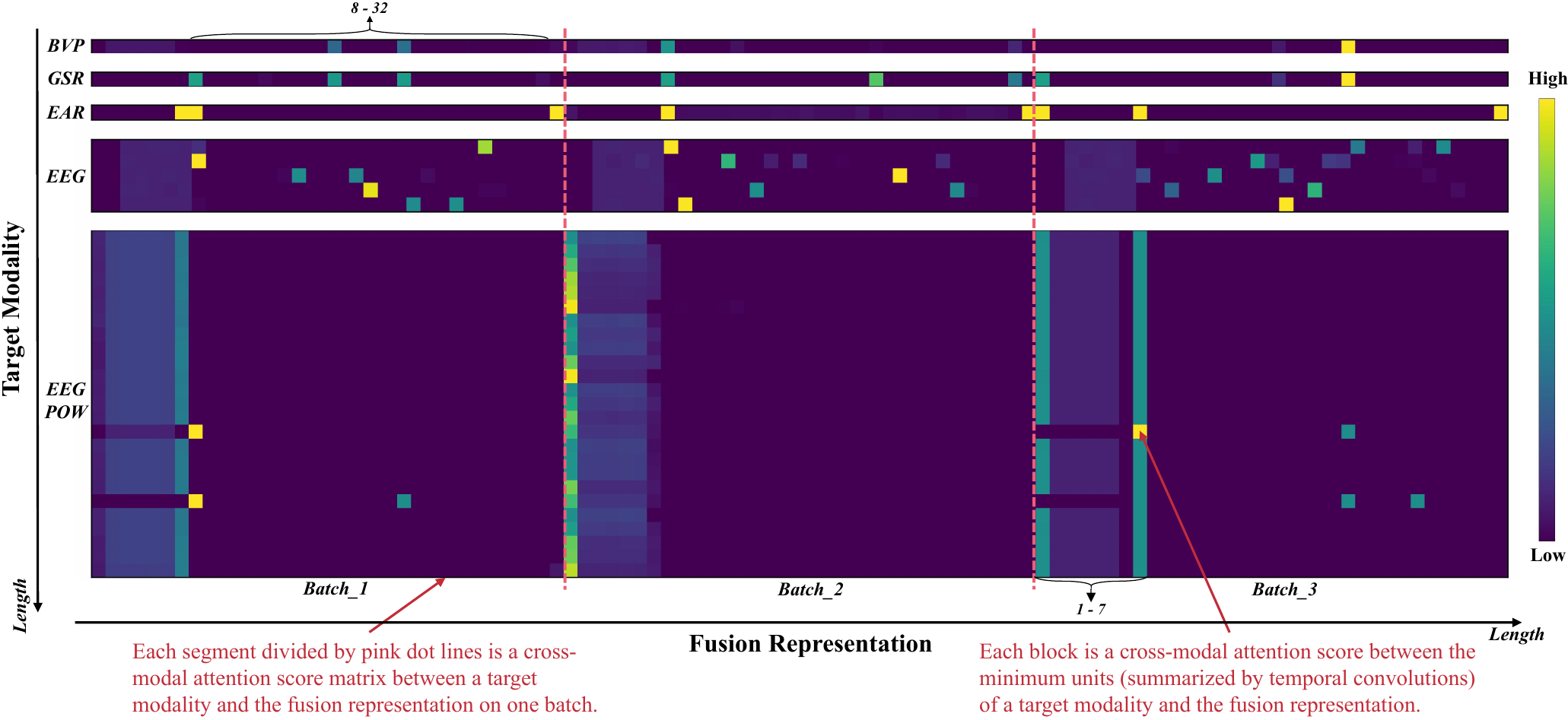}
\caption{Visualization of an example cross-modal attention weight group consisting of learned cross-modal attention matrices at the final layer of each cross-modal transformer within 3 batches during the training on the raw MOCAS dataset. Note that the cross-modal attention score matrix between a target low-level unimodal feature and the low-level fusion representation on one batch has the dimension of ${L_{M_i} \times L_{F}}$, i.e., the length of the target unimodal feature plus that of the fusion representation (BVP, GSR and EAR: $1\times33$; EEG: $5\times33$; EEG\_POW: $25\times33$).  }
\label{cross-v}
\end{figure*}
    
\begin{figure*}[htb]
    \centering
    \subfloat[\label{self-v}]{\includegraphics[width=2.0\columnwidth]{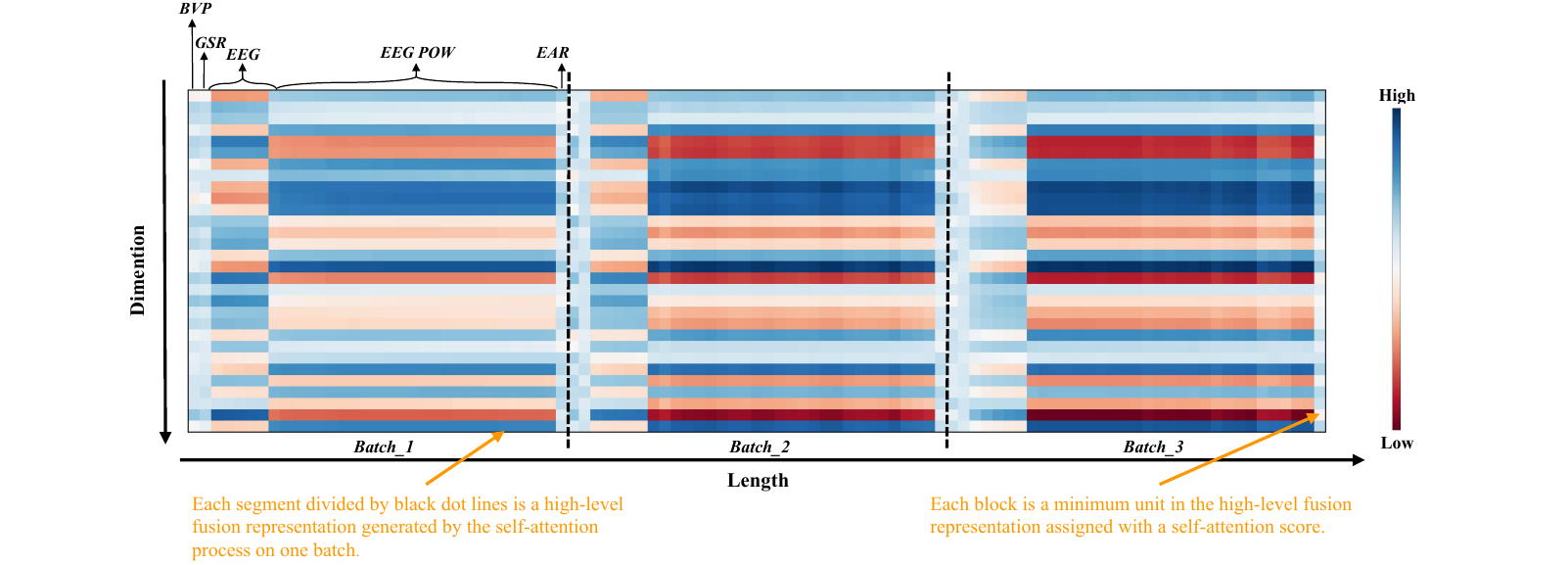}}
    \vspace{-10pt}
    \subfloat[\label{self-v-d}]{\includegraphics[width=2.0\columnwidth]{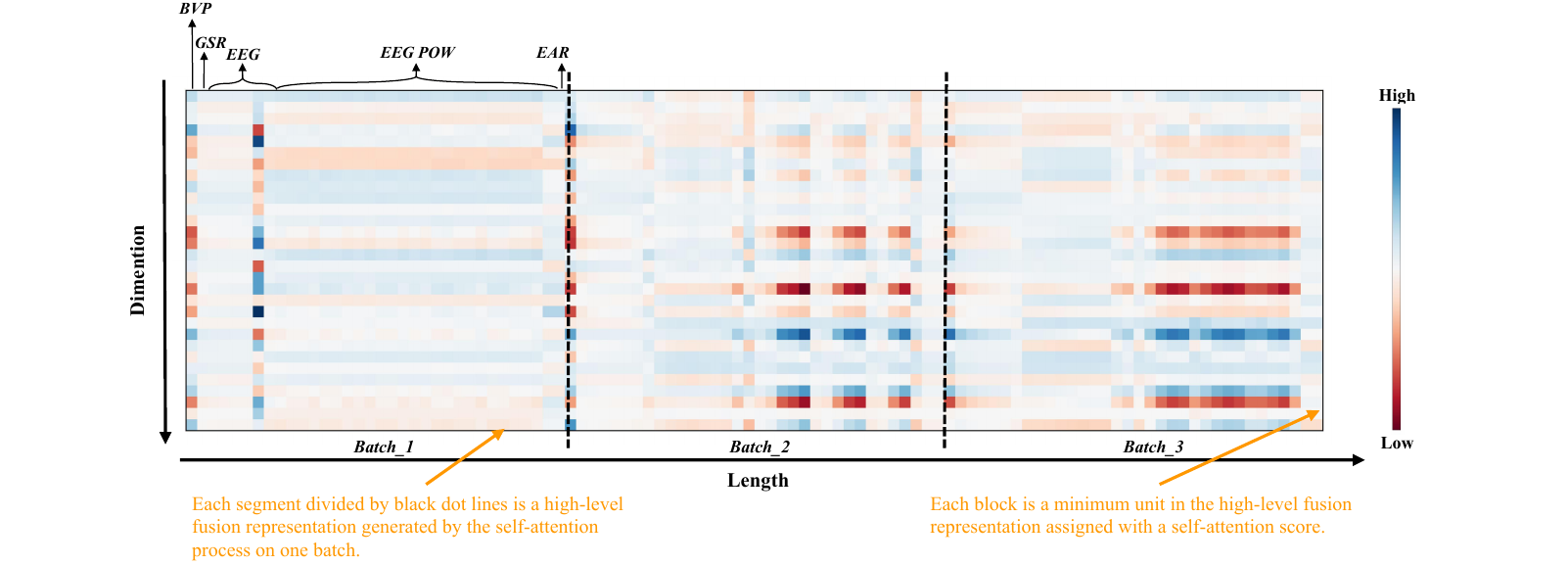}}
    \caption{Visualization of an example high-level fusion representation group generated by the final layer of the self-attention transformer in the (a) \textit{Husformer} and (b) \textit{HusFuse} within 3 batches during the training on the raw MOCAS dataset. Note that the high-level fusion representation produced on one batch has the dimension of ${L_{F} \times D}$, i.e., $33\times30$.}
\end{figure*}

    \textbf{Effectiveness of the self-attention module}
    \begin{itemize}

    \item  Compared with the \textit{HusLSTM} that replaces the self-attention module in the \textit{Husformer} with an LSTM layer, the \textit{Husformer} achieves an absolute improvement on $Acc$ and $F1$. We believe that such an improvement results from the fact that the self-attention mechanism applied in the transformer network \cite{vaswani2017attention} enables the model to capture long-term temporal dependencies by considering the sequence consisting of all reinforced unimodal features as a whole. In contrast, the LSTM processes the sequence element by element, which can suffer from long-dependency issues   \cite{zeyer2019comparison}. Moreover, the self-attention mechanism can adaptively highlight meaningful contextual information while reducing useless one by computing adaptive attention scores at a different position in the sequence, leading to a more effectual global representation of the human state. This result confirms the effectiveness of the self-attention module in the \textit{Husformer}.

\end{itemize}

\subsection{Qualitative Analysis}

To demonstrate how the cross-modal attention and self-attention in the \textit{Husformer} work when learning from multi-modal signals of the human state, we visualize the attention activation for qualitative analysis. Figure \ref{cross-v} shows an example cross-modal attention weight group consisting of learned cross-modal attention matrices at the final layer of each cross-modal transformer within 3 batches during the training on the raw MOCAS dataset. Note that the original cross-modal attention score matrix between a target low-level unimodal feature and the low-level fusion representation on one batch has the dimension of ${L_{M_i} \times L_{F}}$ (see Figure \ref{fig:cm}). We can observe that the cross-modal attention has learned how to attend to positions revealing relevant and meaningful information across the target modality and source modalities embedded in the fusion representation without the requirement of feature alignment. For instance, higher cross-modal attention scores are assigned to some intersections of the BVP and GSR unimodal features and part of the EEG\_POW unimodal features embedded in the later part (8-32) of the fusion representation. This shows that our cross-modal attention can reveal cross-modal contingencies that are inaccessible with manual feature alignment.


Furthermore, we can observe that the learned cross-modal attention is adaptive; i.e., different cross-modal attention patterns are learned between different target modalities and the fusion representation. Moreover, these patterns may differ for the same target modality across different patches. For instance, while the EEG modality is always encouraged to attend to the later section (8-32) of the fusion representation, i.e., the EEG\_POW modality, the intersections assigned with higher cross-modal attention scores vary across different batches. However, despite the above adaptive differences, we can notice that some stable and consistent cross-modal attention patterns exist in different batches of the same target modality. For example, higher cross-modal attention scores are always assigned to the intersections of the EEG\_POW unimodal features and the BVP, GSR and EEG unimodal features embedded in the front positions (1-7) in the fusion representation. These observations over the visualized cross-modal attention score matrices demonstrate that our proposed cross-modal attention module can capture and model an adaptive but relatively consistent and long-term pattern of cross-modal interactions.

Figure \ref{self-v} shows an example of the output of the self-attention transformer in the \textit{Husformer}, namely the high-level fusion representation $Z_{F} \in \mathbb{R}^{L_{F},D}$, within 3 batches during training on the raw MOCAS dataset. We can observe that the self-attention has learned how to prioritize important contextual information for human state recognition and reduce insignificant one by assigning high (blue) and low (red) self-attention scores to different positions in the fusion representation. Also, similar to the cross-modal attention, the learned self-attention is adaptive; i.e., different minimum units in the fusion representation are assigned with self-attention scores in different patterns, especially from the ‘Dimension’ axis. Meanwhile, we can notice that the learned self-attention patterns of the features of the same modality share many similarities across different batches, especially for EEG and EEG\_POW modalities. These observations demonstrate that the self-attention in the \textit{Husformer} can highlight effectual contextual features of human state and diminishes ineffectual ones in the fusion representation with an adaptive while relatively steady pattern.

Meanwhile, Figure \ref{self-v-d} depicts an example of the high-level fusion representation output of the self-attention transformer in the \textit{HusFuse}. Note that the inputs of the self-attention transformers in the \textit{Husformer} and \textit{HusFuse} are concatenated by unimodal features of each modality within the same fragments. The only difference is that the unimodal features in the input of the \textit{Husformer} are reinforced by cross-modal attention transformers while those in the \textit{HusFuse} are not. Comparing the different self-attention patterns assigned to the same fragments of unimodal features in the fusion representation as illustrated in Figure \ref{self-v} and \ref{self-v-d}, we can notice that without the cross-modal attention modelling the cross-modal interactions, the self-attention learned in the \textit{HusFuse} is less efficient. That is, only few of high (blue) and low (red) self-attention scores are assigned to features in the fusion representation, leading to insufficient prominence of critical contextual information and diminishing of unimportant information respectively. We can also notice that the self-attention is less consistent. That is, no stable self-attention patterns are shown on features of the same modality across batches, especially between \textit{Batch\_1} and \textit{Batch\_3}. Such differences demonstrate that the reinforcements for unimodal features from the cross-modal attention in the \textit{Husformer} can help the self-attention highlight critical contextual information in the fusion representation with a more efficient and consistent pattern.

\subsection{Real-world Experiment}
In addition, we applied the \textit{Husformer} to a real-world CCTV monitoring task scenario \cite{jo2023affective}. The \textit{Husformer} effectively predicted the cognitive workload (\textit{low} vs. \textit{medium} vs. \textit{high}) of multiple humans when performing the CCTV monitoring task with a multi-robot system at 100~Hz in real-time. Additionally, by utilizing the predicted objective cognitive load from the \textit{Husformer} model, the overall performance of the human-robot team can be enhanced by adjusting the workload of each human operator accordingly. More details can be found in \cite{jo2023affective}.

\section{Conclusion and Future Work}
\label{sec:conclusion_futureworks}

In this paper, we proposed the \textit{Husformer}, an end-to-end multi-modal transformer framework for recognizing multi-modal human states, including affective states and cognitive load. The \textit{Husformer} fuses modalities with adaptive and sufficient cross-modal interactions, enabling one modality to attend to features of other modalities where strong cross-modal relevance exists. It also adaptively highlights important contextual information in the fusion representation. These two attention mechanisms, operating at the inter-modal and fusion representation levels, enable our model to efficiently learn from multi-modal features, eliminating the need for extensive feature engineering and alignment required in previous works. Our experimental results on four public benchmark multi-modal datasets of human emotion and cognitive load demonstrated the effectiveness of the proposed \textit{Husformer} for general human state recognition, outperforming four other state-of-the-art multi-modal-based baselines and demonstrating enhanced performance over using a single modality. Additionally, our ablation study highlighted the effectiveness of two key components in the \textit{Husformer}: the cross-modal attention and self-attention modules. Through the visualization of attention activation, we demonstrated that the cross-modal attention and self-attention transformers introduced in \textit{Husformer} can respectively model adaptive but relatively consistent and long-term cross-modal interactions of multiple modalities and contextual interactions in the high-level fusion representation. 

In the future, we plan to integrate user sensitivity considerations into the framework to improve its performance and generalization ability. Additionally, we will explore the application of the \textit{Husformer} to more real-world scenarios and investigate its robustness and adaptability in various practical scenarios.



\section*{Acknowledgements}
This material is based upon work supported by the National Science Foundation under Grant No. IIS-1846221. Any opinions, findings, and conclusions or recommendations expressed in this material are those of the author(s) and do not necessarily reflect the views of the National Science Foundation.

\bibliography{main}
\bibliographystyle{IEEEtran}

\newpage
\onecolumn

\appendices
\section{Hyper-parameters of the \textit{Husformer}} \label{appendix:A}

\begin{table*}[h]
    \centering
    \caption{Hyper-parameters of the \textit{Husformer} utilized for each experiment}
    \label{hypara}
    \begin{tabular}{c||c||c||c||c||c||c}
        \hline
        Parameter name & Raw DEAP & Preprocessed DEAP & WESAD & Raw MOCAS & Preprocessed MOCAS & Cogload \\ \hline\hline
        Batch Size                         & 1024     & 1024              & 512        & 64        & 128  & 1024              \\
        Initial Learning Rate              & 2e-3     & 2e-3              & 1e-3      & 1e-3      & 1e-3 & 1e-3              \\
        Optimizer                          & Adam     & Adam              & Adam  & Adam    & Adam      & Adam               \\
        Transformer Hidden Unit Size  & 40        & 40                 & 40     & 40       & 40         & 40  \\
        Crossmodal Attention Heads         & 3        & 3                 & 3            & 5         & 5 & 3                 \\
        Crossmodal Attention Block Dropout & 0.1     & 0.1              & 0.05      & 0.05      & 0.05& 0.1               \\
        Output Dropout & 0.1     & 0.1              & 0.1  & 0.1    & 0.1      & 0.1 \\
        Focal Loss $\alpha_c$ & [0.1,0.1,0.8]     & [0.15,0.05,0.8]              & [0.4,0.3,0.3]  & [0.2,0.1,0.7]    & [0.15,0.15,0.7]      & [0.1,0.1,0.8] \\
        Focal Loss $\gamma$ & 3     & 3              & 2  & 3    & 3      & 2 \\
        Epochs                          & 20       & 20                & 60         & 40        & 40   & 80               \\ \hline
    \end{tabular}
\end{table*}

\section{Classification results using each single modality} \label{appendix:B}


\begin{table*}[h]
\centering
\caption{Best performing classification results of using single
modality on the raw DEAP and preprocessed DEAP Dataset in terms of multi-class average accuracy ($Acc$) and multi-class average F1-score ($F1$) with stand deviations. All best performing results are obtained with the transformer network.}
    \begin{tabular}{c||cc|cc||cc|cc}
    \hline
    Dataset  & \multicolumn{4}{c||}{Raw DEAP}                                                                                              & \multicolumn{4}{c}{Preprocessed DEAP}                                                                                     \\ \hline\hline
    Criteria & \multicolumn{2}{c|}{Valence}                                           & \multicolumn{2}{c||}{Arousal}                       & \multicolumn{2}{c|}{Valence}                                           & \multicolumn{2}{c}{Arousal}                       \\
    Metric   & $Acc(\%)^h$                    & \multicolumn{1}{c|}{$F1(\%)^h$}                     & $Acc(\%)^h$                    & \multicolumn{1}{c||}{$F1(\%)^h$} & $Acc(\%)^h$                    & \multicolumn{1}{c|}{$F1(\%)^h$}                     & $Acc(\%)^h$                    & \multicolumn{1}{c}{$F1(\%)^h$} \\ \hline\hline
    EEG      & 72.80$\pm$2.03 & 72.93$\pm$2.20   & 73.27$\pm$1.72 & 73.71$\pm$1.93  & 84.98$\pm$1.40 & 85.09$\pm$1.42   & 84.86$\pm$0.30 & 84.86$\pm$0.80 \\
    EMG      & 64.66$\pm$2.38 & 64.71$\pm$2.33   & 65.32$\pm$2.24 & 65.66$\pm$1.83  & 76.69$\pm$1.04 & 76.71$\pm$4.02   & 74.09$\pm$0.89 & 74.07$\pm$0.86 \\
    EOG      & 46.99$\pm$1.93 & 45.98$\pm$1.83   & 48.27$\pm$1.57 & 50.97$\pm$1.44  & 62.76$\pm$1.85 & 65.57$\pm$1.45   & 64.42$\pm$0.44 & 65.43$\pm$0.39 \\
    GSR      & 45.18$\pm$2.56 & 45.63$\pm$2.38   & 46.70$\pm$2.33 & 48.07$\pm$2.17  & 61.65$\pm$1.16 & 59.74$\pm$0.97   & 62.75$\pm$0.91 & 57.28$\pm$0.88 \\ \hline
    \end{tabular}
\end{table*}

\begin{table*}[h]
\centering
\caption{Best performing classification results of using single
modality on the raw MOCAS and preprocessed MOCAS Dataset in terms of multi-class average accuracy ($Acc$) and multi-class average F1-score ($F1$) with stand deviations; $^\clubsuit$: classification with Transformer, and $^\spadesuit$: classification with GCN.}
    \begin{tabular}{c||c|c||c|c}
    \hline
    Dataset & \multicolumn{2}{c||}{Raw MOCAS}                                                             & \multicolumn{2}{c}{Preprocessed MOCAS}                                                    \\ \hline\hline
    Metric      & $Acc(\%)^h$                                        & $F1(\%)^h$                                          & $Acc(\%)^h$                                        & $F1(\%)^h$                                         \\ \hline\hline
    EEG         & 34.17$\pm$0.57$^\spadesuit$       & 34.01$\pm$0.53$^\spadesuit$       & 44.80$\pm$0.37$^\clubsuit$        & 46.21$\pm$0.37$^\spadesuit$       \\
    EEG\_POW    & 68.25$\pm$1.31$^\clubsuit$        & 67.87$\pm$1.45$^\clubsuit$        & 84.98$\pm$1.74$^\clubsuit$        & 84.90$\pm$1.72$^\clubsuit$        \\
    GSR         & 33.70$\pm$0.61$^\spadesuit$       & 35.59$\pm$0.64$^\spadesuit$       & 43.44$\pm$0.57$^\clubsuit$        & 46.98$\pm$0.53$^\clubsuit$        \\
    BVP         & 42.55$\pm$1.63$^\spadesuit$       & 43.49$\pm$1.71$^\clubsuit$        & 71.18$\pm$2.03$^\clubsuit$        & 71.14$\pm$2.06$^\clubsuit$        \\
    EAR         & 47.34$\pm$3.12$^\clubsuit$        & 49.29$\pm$2.29$^\clubsuit$        & 51.46$\pm$0.42$^\clubsuit$        & 48.37$\pm$0.45$^\clubsuit$        \\ \hline
    \end{tabular}
\end{table*}

\begin{table*}[h]%
\captionsetup[subfloat]{position=top}
\centering
\caption{Best performing classification results of using single
modality on the (a) WESAD Dataset and (b) CogLoad Dataset in terms of multi-class average accuracy ($Acc$) and multi-class average F1-score ($F1$) with stand deviations. $^\clubsuit$: classification with Transformer; $^\spadesuit$: classification with GCN.}
\subfloat[]{\label{Tab_a}
    \begin{tabular}{|c||c|c|}
    \hline
    Dataset & \multicolumn{2}{c|}{WESAD}                                                        \\ \hline\hline
    Metric      & $Acc(\%)^h$                                        & $F1(\%)^h$                                \\ \hline\hline
    
    EMG         & 52.71$\pm$0.46$^\spadesuit$       & 55.62$\pm$0.54$^\spadesuit$                     \\
    EDA         & 53.85$\pm$0.77$^\clubsuit$        & 56.45$\pm$0.52$^\spadesuit$ \\
    BVP         & 60.75$\pm$0.95$^\clubsuit$        & 61.27$\pm$0.99$^\clubsuit$            \\
    RESP        & 64.09$\pm$1.20$^\clubsuit$        & 65.85$\pm$1.04$^\clubsuit$            \\ \hline
    \end{tabular}
}
\subfloat[]{\label{Tab_b}
    \begin{tabular}{|c||cc|}
    \hline
    Dataset & \multicolumn{2}{c|}{Cogload}     \\ \hline\hline
    Metric      & $Acc(\%)^h$           & $F1(\%)^h$            \\ \hline\hline
    GSR         & 56.45$\pm$0.73$^\clubsuit$    & 57.52$\pm$0.80$^\clubsuit$ \\
    HR          & 30.54$\pm$1.13$^\spadesuit$   & 30.03$\pm$1.21$^\spadesuit$ \\
    RR          & 39.88$\pm$2.32$^\spadesuit$   & 42.58$\pm$2.47$^\clubsuit$ \\
    ACC         & 34.59$\pm$1.81$^\spadesuit$   & 33.35$\pm$1.53$^\spadesuit$ \\ \hline
    \end{tabular}
}\\
\end{table*}

\end{document}